\documentclass[article]{JHEP3}

\usepackage{amsmath,amssymb}
\usepackage[dvips]{graphics}
\usepackage{epsfig}
\usepackage{psfrag,comment}

\newcommand{\be}{\begin{equation}} 
\newcommand{\ee}{\end{equation}}
\newcommand{\bea}{\begin{eqnarray}}
\newcommand{\eea}{\end{eqnarray}}
\newcommand{\bean}{\begin{eqnarray*}}
\newcommand{\eean}{\end{eqnarray*}}
\newcommand{\ba}{\begin{aligned}}
\newcommand{\ea}{\end{aligned}}

\newcommand{\sectiono}[1]{\section{#1}\setcounter{equation}{0}}

\newcommand{\figref}[1]{Fig.~\protect\ref{#1}}

\newcommand{\CC}{{\cal C}}

\newcommand{\CF}{{\cal F}}

\newcommand{\CH}{{\cal H}}
\newcommand{\CI}{{\cal I}}

\newcommand{\CN}{{\cal N}}
\newcommand{\CO}{{\cal O}}
\newcommand{\CP}{{\cal P}}

\def\IZ{{\mathbb Z}}
\def\IR{{\mathbb R}}

\def\IP{{\mathbb P}}

\def\IS{{\mathbb S}}

\newcommand{\tr}{{\rm Tr}}
\newcommand{\re}{{\rm e}}
\newcommand{\ri}{{\rm i}}
\newcommand{\rd}{{\rm d}}

\newdimen\tableauside\tableauside=1.0ex
\newdimen\tableaurule\tableaurule=0.4pt
\newdimen\tableaustep
\def\phantomhrule#1{\hbox{\vbox to0pt{\hrule height\tableaurule width#1\vss}}}
\def\phantomvrule#1{\vbox{\hbox to0pt{\vrule width\tableaurule height#1\hss}}}
\def\sqr{\vbox{%
  \phantomhrule\tableaustep
  \hbox{\phantomvrule\tableaustep\kern\tableaustep\phantomvrule\tableaustep}%
  \hbox{\vbox{\phantomhrule\tableauside}\kern-\tableaurule}}}
\def\squares#1{\hbox{\count0=#1\noindent\loop\sqr
  \advance\count0 by-1 \ifnum\count0>0\repeat}}
\def\tableau#1{\vcenter{\offinterlineskip
  \tableaustep=\tableauside\advance\tableaustep by-\tableaurule
  \kern\normallineskip\hbox
    {\kern\normallineskip\vbox
      {\gettableau#1 0 }%
     \kern\normallineskip\kern\tableaurule}%
  \kern\normallineskip\kern\tableaurule}}
\def\gettableau#1{\ifnum#1=0\let\next=\null\else
\squares{#1}\let\next=\gettableau\fi\next}
\title{Nonperturbative effects and nonperturbative definitions in matrix models and topological strings}

\author{Marcos Mari\~no
\\
D\'epartement de Physique Th\'eorique et Section de Math\'ematiques \\
Universit\'e de Gen\`eve, 
CH--1211 Gen\`eve, Switzerland\\
\\
\email{marcos.marino@math.unige.ch}
}
\abstract{
We develop techniques to compute multi-instanton corrections to the $1/N$ expansion in matrix models 
described by orthogonal polynomials. 
These techniques are based on finding trans-series solutions, i.e. formal solutions with exponentially small corrections, to the 
recursion relations characterizing the free energy. We illustrate this method in the Hermitian, quartic matrix model, and we 
provide a detailed description of the instanton corrections in the Gross--Witten--Wadia (GWW) unitary matrix model. Moreover, we 
use Borel resummation techniques and results from the theory of resurgent functions to relate the formal multi-instanton series 
to the nonperturbative definition of the matrix model. We study this relation in the case of the GWW model and its 
double-scaling limit, providing in this way a nice illustration of various mechanisms connecting the resummation of 
perturbative series to nonperturbative results, like the 
cancellation of nonperturbative ambiguities. Finally, we argue that trans-series solutions are also relevant 
in the context of topological string theory. In particular, we point out that in topological string models with both a matrix model and a large $N$ 
gauge theory description, the nonperturbative, holographic definition involves a sum over the multi-instanton sectors of the matrix model.}


\begin{document}



\sectiono{Introduction}

The study of nonperturbative effects in the $1/N$ expansion of matrix models is of great theoretical and practical importance. First of all, they provide 
a toy model for the nonperturbative aspects of the $1/N$ expansion in more complicated theories. Since matrix models are able to describe both 
noncritical string theories as well as topological string theories on certain backgrounds, they provide a reliable arena for nonperturbative computations in 
string theory. Finally, the critical points of matrix models offer a surprising catalogue of scaling behaviors which are relevant in many physical and mathematical problems (see \cite{deift} for a recent overview), and their nonperturbative aspects should be also important in that context. 

Nonperturbative effects in matrix models were identified long ago in terms of 
eigenvalue tunneling \cite{shenker,david}, and they were subsequently analyzed in the so-called double-scaling limit \cite{gm,ds,bk} that describes conformal field theories coupled to gravity (see, for example, \cite{ezj,david,davidvacua,lvm}). Surprisingly, much less is known 
about these effects away from the critical point. In \cite{msw} various general 
results were obtained for one-instanton amplitudes in one-cut matrix models, and it was explicitly shown in many examples 
that these instantons govern the large order behavior of the $1/N$ expansion. 
A very important motivation for this study was the connection between matrix models and topological string theories discovered in 
\cite{dv} and extended in \cite{mm,bkmp} to toric backgrounds. In some simple cases, the methods of \cite{msw} could be applied in order to obtain 
nonperturbative corrections to the total free energy of topological string theory. 

Another way to compute instanton effects in a matrix model is to consider general, unstable multi-cut configurations. 
This was pointed out in \cite{bde}, and developed more recently in \cite{eynard} in the Hermitian case to obtain 
a formal, universal expansion for the matrix model partition 
function which can be regarded as a multi-instanton expansion. 
Both \cite{msw} and \cite{eynard} are based on the geometric description of the $1/N$ expansion in terms of a spectral curve. 
This formalism is very powerful since it gives universal expressions, but it is not so easy to implement in practice. In some cases it is much 
more convenient to compute the relevant amplitudes in the $1/N$ expansion by using 
the method of orthogonal polynomials, which was introduced and developed in \cite{bessis,biz} and played a key role 
in the analysis of the double--scaling limit. For example, if one has to compute the free energies $F_g$ of the quartic matrix model at 
high genera, the method of orthogonal polynomials will be much more efficient than methods based on the spectral curve. 
One could suspect that the same considerations apply to the calculation of nonperturbative effects. 

One of the purposes of this paper is in fact to develop techniques to compute multi-instanton effects in matrix models by using the formalism of 
orthogonal polynomials. The principle behind these techniques is rather simple, and it can be easily motivated by considering the 
double-scaling limit of the matrix model. In this limit, the recursion relations of orthogonal 
polynomials lead to a differential equation (usually called the string equation) for the specific heat. An asymptotic series 
solution of this equation gives then the perturbative free energy. 
However, one can also compute multi-instanton amplitudes from the string equation by considering a {\it trans-series} solution 
to the differential equation, i.e. a solution involving exponentially small corrections to the asymptotics.  The resulting amplitudes are 
double-scaling limits of the full multi-instanton amplitudes off-criticality. In order to obtain these, one notices that the string equation 
is obtained from a {\it difference} equation, and as we will show in this paper, the trans-series solution to this difference 
equation gives a systematic way to compute the nonperturbative, multi-instanton effects of the full matrix model. 

As a first illustration of this method, we revisit a canonical example, namely the 
quartic matrix model, and we recover and extend the results of \cite{msw} for the one-instanton amplitude. One of the advantages of the techniques 
we develop here is that it they can be also applied to {\it unitary} matrix models, where the geometric techniques 
based on spectral curves have not been developed, and we present a detailed analysis of the instanton corrections in the simplest 
unitary model, namely the Gross--Witten--Wadia model \cite{gw,wadia}. This model is very interesting in many respects: it is the model underlying 
two-dimensional Yang--Mills theory on the lattice, it has a double-scaling limit \cite{ps} which describes the simplest minimal 
superstring theory \cite{kms,ss}, and it plays a role in the description of Yang--Mills theory on $\IS^3 \times \IS^1$ (see \cite{abw} for a recent 
study and references to previous work). 

The techniques we develop give the multi-instanton corrections as {\it formal} power series in two small parameters, namely 
$1/N$ and $\re^{-N}$ (this is indeed typical of a quantum-mechanical computation involving instanton effects, like the 
WKB method). Once one has determined these formal, nonperturbative corrections, a natural question is: how is the resulting 
series related to the nonperturbative definition of the model? The second purpose of this paper is to clarify this relation. 
It turns out that, in the cases we will be interested in, 
the trans-series can be resummed with Borel resummation techniques, and one obtains in the end a one-parameter family of 
functions which can be regarded as nonperturbative completions of the theory. The fact that formal solutions to the relevant equations come in families is 
very well known in the case of double-scaled matrix models, but we use results in the theory of exponential asymptotics and the theory of resurgent functions 
to construct actual (convergent) solutions both at criticality and off-criticality. 

We illustrate these ideas in the case of the GWW model and its double-scaling limit. The $1/N$ expansion 
of this model, in the weak coupling regime, has the properties typical of perturbative expansions in realistic 
quantum field theories: the expansion is not Borel summable, yet the model has a unique nonperturbative definition in terms of the 
original unitary integral. Therefore, this is an excellent laboratory to explore the analyticity properties of the $1/N$ 
expansion. We show that the $1/N$ series can be resummed in such a way that the ambiguities coming from the Borel 
resummation cancel against nonperturbative instanton effects, providing a nice illustration of the cancellation mechanism much discussed in 
renormalon physics \cite{davidren,grunberg} and in some quantum-mechanical problems \cite{zjj}. In our case, as in \cite{zjj}, this cancellation 
is a consequence of the ''resurgence" properties of trans-series expansions \cite{ecalle, approche}. The resummation process gives a one-parameter family of solutions 
which include in a crucial way multi-instanton corrections. We show that, for a particular (and rather natural) choice of the parameter, this solution is the semiclassical expansion of the true nonperturbative answer, and multi-instanton corrections are crucial in order to reproduce the exact, nonperturbative value of the different physical quantities. 

Our results for the double-scaling limit of the GWW model have a clear interpretation in terms of 
minimal superstring theory. 
As in the case of bosonic minimal string theories, the minimal superstring has ZZ 
branes which should correspond to 
eigenvalue instantons. We interpret the sum over multi-instantons that arises naturally in the 
trans-series solution 
as a sum over ZZ brane backgrounds, i.e. over sub-leading saddles of the theory. 
It was pointed out in \cite{mmss} that the exact, nonperturbative answer 
for the free energy in noncritical string theories should include indeed a sum over all of these backgrounds. This is 
precisely what we obtain here. The general lesson of our analysis is that, to make sense of this sum, 
one has to be careful about various subtleties. These include Borel resummation of the various asymptotic expansions, 
the choice of nonperturbative parameter, 
the cancellation of non-perturbative ambiguities, and the reality conditions of the final solution. 

As we mentioned before, an important motivation for the study of nonperturbative effects in matrix models is the connection to topological string theory. It is natural to conjecture that the full topological string theory partition function is a trans-series expansion which includes 
nonperturbative multi-instanton effects. This was already suggested in \cite{mmss} following the analogy with noncritical strings, and made more concrete in \cite{msw}. 
In some simple topological string theories with a dual matrix model description, 
the one-instanton sector was studied in \cite{msw} by using their geometric description in terms of spectral curves/mirror symmetry. In this paper we 
give two more pieces of evidence for this conjecture. We first revisit one of the examples of \cite{msw}, namely Hurwitz theory, 
which can be regarded as a toy model of topological string theory. The free energy 
of this theory is described by a difference equation of the Toda type \cite{p}, therefore one can find a trans-series solution of this equation and 
we verify that this solution reproduces the one-instanton effects computed in \cite{msw}. Second, we point out 
that in more complicated topological string models, with both a matrix model and a holographic, large $N$ gauge theory description, 
the nonperturbative, holographic definition tells us that the full partition function involves indeed a sum over instanton sectors in the matrix model.  It has been pointed 
out by Eynard \cite{eynard} 
that such sums should be background independent, since all the backgrounds are summed over, and this opens the possibility that 
a proper understanding of such trans-series solutions will lead to background independent topological string models. 

The organization of this paper is as follows. In section 2 we review some basic results on matrix models in the formalism of orthogonal polynomials. In section 3 
we explain how to obtain multi-instanton amplitudes in matrix models by constructing trans-series solutions to the relevant difference equations. We illustrate the method 
for the quartic matrix model. Section 4 is devoted to a detailed analysis of instanton effects in the GWW unitary matrix model, focusing on the weakly coupled phase. In section 
5 we address the issue of how to relate the formal multi-instanton expansions to the nonperturbative definition of the matrix model (when available). We introduce 
ideas and techniques from Borel resummation and the theory of resurgence, and we perform a very detailed analysis of this issue in the GWW model and its double-scaling 
limit, described by Painlev\'e II. Section 6 uses the framework developed in this paper to analyze topological string theory, building on the results of \cite{msw}. Finally, 
section 7 states our conclusions as well as some open problems.


\sectiono{Matrix models and orthogonal polynomials} 
In this section we will review some elementary aspects of the method of orthogonal polynomials as 
applied to the calculation of the $1/N$ expansion of a matrix model. 
Many of the key formulae that we will need are common to both Hermitian and unitary matrix models, 
but for concreteness we will start discussing the Hermitian case. The method of orthogonal polynomials for Hermitian matrix models was discovered in \cite{bessis, biz}, and useful reviews can be 
found in \cite{dfgzj,mmleshouches}. We will follow the conventions of this last reference. 

We will consider gauged, Hermitian matrix models defined by the partition function
\be
Z={1\over {\rm vol}(U(N))} \int \rd M \re^{-{1\over g_s} \tr \, V(M)}, 
\ee
where $V(M)$ is the potential. A standard argument reduces this integral to an integral over 
eigenvalues
\be
\label{partitionz}
Z={1 \over N!} \int \prod_{i=1}^N {\rd\lambda_i \over 2 \pi} \, \Delta^2(\lambda) \re^{-{1\over g_s} \sum_{i=1}^N 
V(\lambda_i)},
\ee
where
\be
\Delta(\lambda) =\prod_{i<j}(\lambda_i-\lambda_j)
\ee
is the Vandermonde determinant. 
If we regard
\be
\rd\mu = \re^{-{1\over g_s} V(\lambda)} {\rd\lambda \over 2 \pi}
\ee
as a measure
in $\IR$, one can introduce orthogonal polynomials $p_n(\lambda)$ defined by
\be
\int \rd\mu \, p_n(\lambda) p_m(\lambda)  = h_n \delta_{nm},\quad n\ge 0,
\label{ortho}
\ee
where $p_n(\lambda)$ are normalized by requiring the behavior $p_n(\lambda)=\lambda^n +\cdots$. 
One then easily finds, 
\be
\label{parth}
Z =\prod_{i=0}^{N-1} h_i = h_0^N \prod_{i=1}^N r_i^{N-i},
\ee
where we have introduced the coefficients
\be
\label{rcoeff}
r_k= {h_k \over h_{k-1}}, \qquad k\ge 1
\ee
which appear in the recursion relations for the $p_n(\lambda)$, 
\be
\label{recurs}
(\lambda + s_n ) p_n(\lambda) = p_{n+1}(\lambda) + r_n p_{n-1}(\lambda).
\ee

It will be useful to normalize the results by considering the 
Gaussian matrix model, 
\be
Z_G={1\over {\rm vol}(U(N))} \int \rd M \re^{-{1\over 2g_s} \tr \, M^2}, 
\ee
i.e. we will be interested in computing the normalized free energy
\be
F=\log Z -\log Z_G. 
\ee
This free energy has an asymptotic expansion around $g_s=0$ of the form 
\be
\label{largeNf}
F(t,g_s) =\sum_{g=0}^{\infty} F_g(t) g_s^{2g-2}
\ee
where $t$ is the 't Hooft parameter 
\be
t=g_s N. 
\ee
Since we keep $t$ fixed, (\ref{largeNf}) is also a large $N$ expansion in powers 
of $1/N^2$. The standard procedure to compute this asymptotic expansion by using orthogonal polynomials 
goes as follows. We have an exact formula for finite $N$, 
\be
g_s^2 F= {t^2 \over N} \log {h_0\over h_0^G} + {t^2 
\over N}  \sum_{k=1}^N \biggl( 1-{k\over N} \biggr)
\log {r_k \over k g_s},
\label{allf}
\ee
where $h_0^G$ is the coefficient $h_0$ for the Gaussian model. 
In order to proceed, we introduce a continuous variable as $N \rightarrow \infty$, 
\be
\label{continuumvar}
 g_s k \rightarrow z, \qquad 0\le z \le t, 
\ee
and we assume that in this continuum, $N \rightarrow \infty$ limit, $r_k$ becomes a function of $z$ and $g_s$, 
\be
r_k  \rightarrow R (z,g_s) 
\ee
It will be useful to consider the function 
\be
\Xi(z,g_s) = {R(z,g_s) \over z}
\ee
which can be regarded as the continuum limit of $r_k/(k g_s)$. It is easy to see that, for polynomial potentials of the form
\be
V(M) = {1\over 2}M^2 +\cdots
\ee
one has $r_k \sim k g_s +\cdots$, therefore the function $\log (r_k/(k g_s))$ is regular at $k=0$ and we 
can use the standard Euler--Maclaurin summation formula to evaluate (\ref{allf}). One then obtains \cite{bessis,biz}:
\be
\label{ofgex}
\ba
g_s^2 F &= \int_0^t \rd z\,  (t-z) \log \Xi (z)+  \sum_{p=1}^{\infty} g_s^{2p}\, \, {B_{2p} \over (2p)!}\, \frac{\rd ^{2p-1}}{\rd z^{2p-1}} \biggl[ \left( t-z \right) \log \Xi(z,g_s) \biggr] \bigg|_{z=0}^{z=t} \\
&+ {t g_s \over 2 } \biggl[ 2 \log {h_0 \over h_0^{\rm G}} - \log \Xi (0,g_s)\biggr].
\ea
\ee

We will rephrase (\ref{ofgex}) in a more convenient way. A small calculation shows that
\be
g_s^2 {\partial^2 F\over \partial t^2} = \log \Xi(t) - \sum_{p=1}^{\infty} g_s^{2p}\, \, {B_{2p} \over (2p) (2p-2)!}\, \frac{\rd ^{2p}}{\rd t^{2p}}  \log \Xi(t,g_s). 
\ee
We now use the fact that 
\be
z^2 {\rm csch}^2 (z) =1-\sum_{k=1}^{\infty} {2^{2k} B_{2k} \over (2k) (2k-2)!} z^{2k}, 
\ee
to write the above equation as
\be
4  \sinh^2  \biggl({g_s \over 2}{\rd \over \rd t}\biggr) F(t) = \log \Xi. 
\ee
The first member can be written as a difference operator, therefore
\be
\label{diffeq}
F(t+g_s) + F(t-g_s)-2 F(t) =\log \Xi, 
\ee
or equivalently,
\be
\label{todalike}
\exp \Bigl[ F(t+g_s) + F(t-g_s)-2 F(t) \Bigr]=\Xi.
\ee
A shorter way to derive this equation is simply to start from the identity 
\be
{Z_{N+1} Z_{N-1} \over Z_N^2} =r_N, 
\ee
where $Z_N$ is the partition function (\ref{parth}) at rank $N$, and consider its continuum limit. 
Notice that, written in the form (\ref{todalike}), 
the equation determining the free energy involves the standard difference operator of the Toda lattice. 
This is related to the fact that the free energy of a polynomial matrix model is a solution to the Toda hierarchy \cite{morozov}. 

In order to compute the $g_s$ expansion of the free energy (\ref{largeNf}), one finds first an expansion for $R(z,g_s)$ of the form 
\be
\label{zeroex}
R^{(0)}(z,g_s) = \sum_{s=0}^{\infty} g_s^{2s} R_{0,2s}(z).
\ee
Once this expansion is plugged in $\Xi(z,g_s)$ and then in (\ref{ofgex}), the expansion (\ref{largeNf}) follows. 
In order to obtain (\ref{zeroex}) one has to use the so-called 
pre-string equation. This is a difference 
equation for $R(z,g_s)$ which can be derived as the continuum limit of the recursion relations obeyed by the coefficients (\ref{rcoeff}). 
The pre-string equation can be explicitly written for any polynomial potential \cite{biz,dfgzj}. For example, in the 
case of the quartic matrix model with potential 
\be
\label{quarticpot}
V(M) ={1\over 2} M^2 -{\lambda \over 48} M^4, 
\ee
the difference equation for $R(z,g_s)$ reads as
\be
\label{diff}
R(z,g_s)\Bigl\{ 1 -{\lambda\over 12} (R(z,g_s) + R(z+g_s,g_s) + R(z-g_s,g_s) \Bigr\}=z. 
\ee
This type of difference equations have a solution of the form (\ref{zeroex}), and they determine $R_{0,s}(z)$ in terms of the $R_{0,s'}(z)$, $s'<s$. 
When this solution is plugged in (\ref{diffeq}), one obtains the perturbative expansion of the total free energy in powers of $g_s$, 
which is the standard $1/N$ expansion of the matrix model \cite{bessis,biz}. 

The formalism of orthogonal polynomials for unitary matrix models is very similar (see, for example, \cite{ps}). We consider unitary matrix models of the form
\be
\label{uniz}
Z=\int \rd U \, \re^{{1\over g_s}  V(U)},
\ee
where $U$ is a unitary matrix and the potential $V(U)$ has the structure
\be
V(U)=\sum_l \Bigl( g_l {\rm tr}\,  U^l + {\overline g}_l  {\rm tr}\, U^{\dagger l} \Bigr), \qquad g_l={1\over 2 l}(\beta_l - \ri \gamma_l).
\ee
We can write the partition function in terms of the 
eigenvalues of $U$, $\phi_i \in [-\pi, \pi]$:
\be
Z=\int \prod_i \rd\phi_i \prod_{i<j} 4 \sin^2 \biggl( {\phi_i - \phi_j \over 2} \biggr) {\rm e}^{{1\over g_s} \sum_{i=1}^N  V(\phi_i)},
\ee
where
\be
\label{action}
V(\phi)=\sum_l \biggl(  {\beta_l \over l} \cos\, l\phi + {\gamma_l \over l} \sin\, l \phi \biggr).
\ee
If we introduce the measure
\be
\rd\mu ={1\over 2\pi \ri} {\rd z\over z} \re^{{1\over g_s}  V(z)},
\ee
the orthogonal, monic polynomials
\be
p_n(z)=z^n+ \cdots
\ee
satisfy
\be
\oint  \rd\mu \, p_n(z) p_m(z^{-1}) =h_n \delta_{nm}.
\ee
as well as the the recursion relation
\be
\label{fnrec}
p_{n+1}(z)=z p_n(z) + f_n z^n p_n(z^{-1}),
\ee
and one easily shows that
\be
\label{hf}
{h_{n+1}\over h_n}=1-f_n^2.
\ee
As in the Hermitian case, we introduce the quantities
\be
r_n={h_n\over h_{n-1}}.
\ee
In terms of these, the partition function of the unitary matrix model is given again by the 
formula (\ref{parth}). Normalizing by the Hermitian, Gaussian matrix model, and introducing the 
continuum limit (\ref{continuumvar}) for the $r_n$, we find the same formalism describing both the unitary and the Hermitian matrix model. 
In both cases the key ingredient is to derive explicit expressions for 
the function $R(z,g_s)$. In the unitary case these are also obtained by solving a difference equation, 
which will depend on the particular model one is considering. A particularly important example, 
the Gross--Witten--Wadia (GWW) model \cite{gw,wadia}, will be studied in section 4. 

\sectiono{Multi-instanton corrections} 

In this section we develop techniques to compute multi-instanton corrections in matrix models 
with the method of orthogonal polynomials. We first describe the general structure of the method, and then, as an example, we 
analyze in some detail the Hermitian matrix model with an even, quartic potential. 

\subsection{Multi-instantons and matrix models}

Before presenting our method we will provide a general framework for instanton calculus in matrix models, referring to for example 
\cite{david,davidvacua,lvm,msw,bde,eynard} for more details. In this section we will consider Hermitian matrix models with polynomial potentials. 

The expansion of the matrix model partition function $Z$ in even powers of $g_s$ is in fact 
an asymptotic expansion of $Z$ around a saddle point of the matrix integral. 
These saddle points are characterized by a distribution of matrix eigenvalues $\rho(\lambda)$. 
For example, in the so-called one-cut case, all the eigenvalues sit on the same interval, which is located around a minimum of the potential. 
Let us assume that the potential of the matrix model has $d$ different extrema $x_1, \cdots, x_{d}$. Then, the most general saddle-point is a configuration 
in which the $N$ eigenvalues split into $d$ sets of $N_k$ eigenvalues, $k=1, \cdots, d$. Let us denote each of these $d$ sets by 
\be
\{ \lambda^{(k)}_{i_k}\}_{i_k=1, \cdots, N_k}, \quad k=1, \cdots, d. 
\ee
The eigenvalues in the $k$-th set sit in an interval or arc $\CI_k$ around the $k$-th extremum. Along this interval, the effective potential
\be
V_{\rm eff}(\lambda) =V(\lambda) -t \int \rd \lambda' \rho(\lambda') \log|\lambda -\lambda'| 
\ee
is constant. It is possible to choose $d$ integration contours $\CC_k$ in the complex plane, 
$k=1, \cdots, d$, going to infinity in directions where the integrand decays exponentially, and in such a way 
that each of them passes through exactly one of the $d$ critical points (see for example \cite{fr}). The resulting matrix integral is convergent and can be written as 
\be
\label{genz}
Z(N_1, \cdots, N_d)={1\over N_1! \cdots N_d!} \int_{\lambda^{(1)}_{i_1} \in \CC_1} \cdots \int_{\lambda^{(d)}_{i_d} \in \CC_d}  
\prod_{i=1}^N {\rd\lambda_i \over 2 \pi}\, \Delta^2(\lambda) \re^{-{1\over g_s} \sum_{i=1}^N 
V(\lambda_i)}.
\ee
Of course, when the integrand is written out in detail, it splits into $d$ sets of eigenvalues which interact 
among them through the Vandermonde determinant (see for example \cite{kmt}). If one now regards (\ref{genz}) as the matrix integral 
in a topological sector characterized by the numbers $N_1, \cdots, N_d$, it is natural to consider the general partition function 
\cite{david,davidvacua,bde,eynard} 
\be
\label{sumz}
Z =\sum_{N_1 +\cdots +N_d=N} \zeta_1^{N_1} \cdots \zeta_d^{N_d} Z(N_1, \cdots, N_d).
\ee
The coefficients  $\zeta_k$ can be regarded as $\theta$ parameters which lead to different $\theta$ vacua \cite{davidvacua}\footnote{This type of structure has been argued to be relevant as well for general QFT path integrals \cite{garcia,guralnik}.}. Notice that one can fix the overall normalization by setting one of the $\zeta_k$'s to $1$, for example. The sum (\ref{sumz}) can be also regarded as a matrix integral where the $N$ eigenvalues are integrated along the contour
\be
\CC =\sum_{k=1}^d \zeta_k \CC_k, 
\ee
therefore the $\theta$ parameters give the relative weight of the different contours $\CC_k$ \cite{davidvacua,eynard}. 

Among all configurations characterized by the fillings $(N_1, \cdots, N_d)$, the most stable one occurs when all the eigenvalues sit at the minimum of the potential, which we will take to be $x_1$ (we assume for simplicity that this minimum is unique). We will regard the resulting configuration 
\be
\label{refcon}
(N,0, \cdots, 0)
\ee
as the reference configuration for the system. This corresponds to a one-cut solution of the matrix model. 
It is then easy to see that the other terms in the sum (\ref{sumz}), with general filling numbers $N_i$, 
are exponentially suppressed with respect to the reference configuration (\ref{refcon}), with a weight of the form 
\be
\exp \biggl\{ -{1\over g_s} \sum_{i=2}^d N_i (V_{\rm eff}(x_i) -V_{\rm eff}(x_1)) \biggr\}
\ee
and can then be regarded as {\it instanton} configurations.  

\FIGURE[ht]{
\leavevmode
\centering
\epsfysize=3.5cm
 \hspace{3cm} \epsfbox{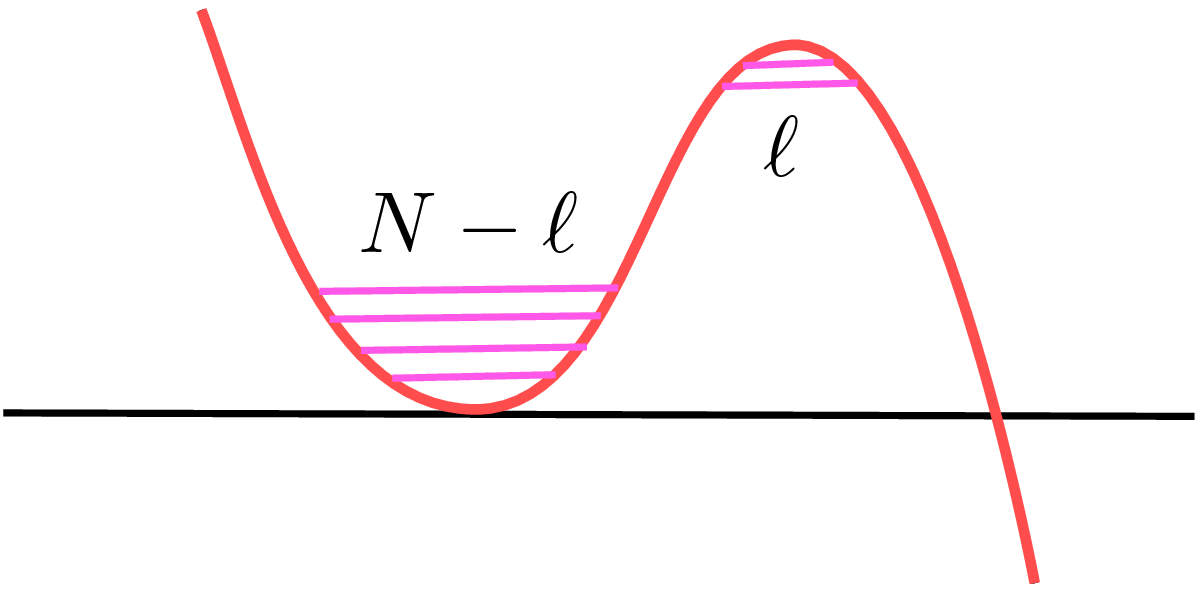}   \hspace{3cm}
\caption{An $\ell$-instanton configuration in a matrix model with a cubic potential. $N-\ell$ eigenvalues sit at 
the minimum, while $\ell$ eigenvalues sit at the maximum.}
\label{cubicuts}
}

An example of such a situation is the cubic matrix model, where the potential has 
two different extrema (a minimum and a maximum). In the reference configuration 
all the eigenvalues of the matrix sit near the minimum. This is the standard one-cut solution described by the 
method of orthogonal polynomials, and gives $Z(N, 0)$. The configuration in which $\ell$ eigenvalues sit 
near the maximum is an $\ell$-instanton of the cubic matrix model, and it gives the partition 
function $Z(N-\ell,\ell)$ (see \figref{cubicuts}). After summing over all topological sectors, and fixing the normalization by setting $\zeta_1=1$, $\zeta_2=\zeta$, we obtain the 
partition function 
\be
Z =\sum_{\ell=0}^{N} \zeta^\ell Z(N-\ell,\ell),
\ee
which in the large $N$ limit becomes
\be
\label{znl}
Z =\sum_{\ell=0}^{\infty} \zeta^\ell Z^{(\ell)}(t,g_s).
\ee

The method of orthogonal polynomials, as presented in \cite{bessis,biz} and summarized in the previous section, 
gives tools to compute the asymptotic $g_s$ expansion of the one-cut answer 
$Z^{(0)}(t,g_s)$, and it is natural to ask if we can use it to calculate the instanton corrections to $Z^{(0)}(t,g_s)$, i.e. the partition functions 
$Z^{(\ell)}(t,g_s)$, $\ell\not=0$. Each of these should have an asymptotic expansion in $g_s$, which corresponds physically to the 
perturbative $g_s$ expansion around the instanton configuration. 
In this paper we will develop techniques to do that. We will focus on the situation that we have just considered, namely, a 
matrix integral where the reference configuration is the one-cut solution, and the instantons are obtained by moving a small number of eigenvalues from the minimum to another saddle point. 

It is worth mentioning that (\ref{genz}) is nothing but a partition function for a multicut matrix model, and one should be able to 
evaluate a generic instanton configuration as a particular 
case of the multicut theory. This was pointed out in \cite{bde} and developed recently in more detail in \cite{eynard}, where general expressions were obtained for the formal expansion of (\ref{genz}) and (\ref{sumz}) in the Hermitian case, and in the framework of the full saddle-point $1/N$ expansion presented in \cite{eo}. Here we 
are interested in computing this formal expansion in concrete models described by orthogonal polynomials, and to work out the precise relation 
between the formal expansion and the original integral. Some aspects of the relation between multi-instantons and multicut models will be studied in \cite{mswta}. 

\subsection{Multi-instantons and trans-series}
In order to understand our approach to the calculation of nonperturbative effects in matrix models 
off-criticality, it is useful to look first at this problem in the double scaling limit (see \cite{dfgzj} for a review of the double-scaling 
limit of matrix models). For simplicity we will consider the example 
of pure 2d gravity, which can be obtained for example from the quartic or the cubic Hermitian matrix models. 
In this limit, the total free energy $F_{\rm ds}(\kappa)$, as a function of the cosmological constant 
$\kappa$, is described by a set of two equations. The first one relates $F_{\rm ds}$ to the specific heat $u$, and reads
\be
\label{dsf}
F_{\rm ds}''(\kappa)=-u(\kappa). 
\ee
This is in fact the double-scaling limit of (\ref{diffeq}). The second equation  
is a differential equation for $u$. For 2d gravity one obtains Painlev\'e I (we use 
the normalization appropriated for the quartic matrix model)
\be
\label{p1}
-{1\over 3} u'' +u^2=\kappa. 
\ee
The asymptotic solution of this equation which goes like $u\sim {\sqrt {\kappa}}$ as $\kappa \rightarrow \infty$,  
\be
u^{(0)} (\kappa) ={\sqrt {\kappa}}  \sum_{g=0}^{\infty} u_{0,g} \kappa^{-5g/2}
\ee
describes the perturbative free energy. However, one can find a one-parameter family of solutions to 
(\ref{p1}) which includes exponentially suppressed terms as $\kappa\rightarrow \infty$:
\be
\label{transseries}
u(\kappa) = \sum_{\ell=0}^{\infty}C^{\ell} u^{(\ell)}(\kappa)=
 {\sqrt {\kappa}}  \sum_{\ell=0}^{\infty} C^{\ell} \kappa^{-{5 \ell \over 8}} \re^{-\ell A \kappa^{5/4}} \epsilon^{(\ell)}(\kappa), 
\ee
where $C$ is a parameter, the constant $A$ has the value
\be
\label{aaction}
A={4 {\sqrt 6}\over 5}
\ee
and 
\be
\label{el}
\epsilon^{(\ell)}(\kappa)=\sum_{n=0}^{\infty} u_{\ell,n+1} \kappa^{-5n/4}
\ee
are asymptotic series. Since we have introduced an arbitrary constant $C$ in (\ref{transseries}), we can 
normalize the solution such that $u_{1,0}=1$. 

These types of solutions to differential equations are called {\it trans-series}, and are the 
central object in the theory of exponential asymptotics. By plugging (\ref{transseries}) in 
(\ref{dsf}), we obtain a similar trans-series expansion for the free energy, 
\be
F_{\rm ds}(\kappa)= \sum_{\ell=0}^{\infty}C^{\ell} F^{(\ell)}_{\rm ds} (\kappa)= 
\sum_{\ell=0}^{\infty} C^{\ell} \kappa^{-{5 \ell \over 8}} \re^{-\ell A \kappa^{5/4}} \varphi^{(\ell)}_{\rm ds} (\kappa), 
\ee
where $F_{\rm ds}^{(0)}$ is the perturbative free energy. The exponentially suppressed corrections in the 
trans-series expansion can not be seen in a perturbative expansion around $\kappa=\infty$ and their origin 
is a nonperturbative effect. 

Several remarks can be made concerning these solutions. The first one is that 
(\ref{transseries}) is an expansion in two small parameters, namely 
\be
x=\kappa^{-5/4}, \qquad \xi =\kappa^{-{5  \over 8}} \re^{- A \kappa^{5/4}}.
\ee
Usually one first expands in $\xi$ in order to extract the $\ell$-th term $u^{(\ell)}$ 
in the trans-series, and then one expands in $x$ in order 
to obtain the asymptotic expansion of this term, but in some cases it is useful to first expand in $x$ \cite{costincostin}. 
The second remark is that (\ref{transseries}) is a one-parameter family, parametrized by $C$. This corresponds to the nonperturbative 
ambiguity plaguing these problems. The third remark is that, as for any asymptotic expansion, the trans-series 
solution is only valid in a sector of the complex plane, and as we go from one sector to another and  
cross a Stokes line the asymptotics will change. The Stokes line in the example above is the positive real axis ${\rm arg}(\kappa)=0$. 
However, the difference between the two asymptotic solutions 
as we cross a Stokes line will be a shift in the parameter $C$ appearing in (\ref{transseries}), 
\be
C \rightarrow C+S, 
\ee
where $S$ is sometimes called the Stokes multiplier. There are many ways to obtain $S$ in the case of 
Painlev\'e I. One can for example deduce it from a matrix model calculation, as first done by David in \cite{davidvacua}, 
or one can derive it rigorously in the framework of isomonodromy deformations, see \cite{fik} and the comprehensive book \cite{painlevet}. 

What is the nonperturbative origin of the exponentially suppressed terms in the trans-series expansion? In the case of the expansion describing 
the double-scaling limit of matrix models, 
these effects are due to multi-instantons. The $\ell$-instanton 
correction $F^{(\ell)}_{\rm ds} (\kappa)$, which is obtained from the full trans-series 
solution for $u(\kappa)$, can be computed by taking the double-scaling limit of an $\ell$-instanton configuration of the appropriate matrix model partition 
function. For Painlev\'e I, one can take for example the cubic matrix model, and 
the free energy $F^{(\ell)}_{\rm ds} (\kappa)$ can be computed as the double scaling limit of $\log Z^{(\ell)}(t,g_s)$ appearing in (\ref{znl}). 
For $\ell=1$, the instanton origin of the exponentially suppressed corrections to Painlev\'e I has been verified 
by a direct calculation in \cite{david,davidvacua,lvm,msw}, in the context of the saddle-point solution to Hermitian matrix models. 

Of course, it is much more efficient to compute 
multi-instanton effects in the double-scaling limit by using the trans-series solution to the 
string equation. It is then natural to ask if there is such a direct way of computing multi-instanton effects in the full matrix model, away from the critical point. 
Indeed, it is very easy to lift the computation in terms of the string equation to the original matrix integral. 
Recall that the $1/N$ or $g_s$ expansion of the full matrix model is described, in the formalism of orthogonal polynomials, 
by a function $R(z,g_s)$ which satisfies a difference 
equation. In the double-scaling limit, $R(z,g_s)$ leads to the specific heat $u$, and the difference equation 
satisfied by $R(z,g_s)$ leads to the differential equation satisfied by $u$ (the string equation). 
Difference equations, just like differential equations, also admit 
trans-series solutions, and one could suspect that the trans-series solution to the difference equation governing 
$R(z,g_s)$ encodes the multi-instanton amplitudes of the full matrix model. 
To obtain the trans-series solutions, we consider a more general ansatz than (\ref{zeroex}), 
\be
\label{transr}
R(z,g_s) = \sum_{\ell=0}^{\infty} C^{\ell} R^{(\ell)}(z,g_s), 
\ee
where $R^{(0)}(z,g_s)$ is given by (\ref{zeroex}), and for $\ell\ge1$ we have
\be
\label{kpert}
 R^{(\ell)}(z,g_s)=  \re^{-\ell A(z)/g_s} R_{\ell,1}(z) \Bigl( 1+\sum_{n=1}^{\infty} g_s^{n} R_{\ell,n+1}(z) \Bigr),\qquad \ell \ge 1.
 \ee
 Once this ansatz is plugged in the difference equation for $R(z,g_s)$, one obtains a recursive system of equations for the different quantities involved. The 
 quantity $A(z)$, which is a parameter-dependent instanton action, is determined by an equation of the form 
 \be
 A'(z) =f(R_{0,0}(z)), 
 \ee
 where $f$ is a function fixed by the difference equation. For $\ell=1, n>0$, one 
 obtains an equation which determines 
\be
{\rd R_{1,n} (z) \over \rd z}
\ee
in terms of $R_{1,n'}(z)$ with $n'<n$. For $n=1$, we have a differential equation for the logarithmic derivative, i.e. for
\be
{1\over R_{1,1} (z)} {\rd R_{1,1} (z) \over \rd z}. 
\ee
The integration constant for $R_{1,1}(z)$ can be reabsorbed in the parameter $C$, and for $A(z)$ and the $R_{1,n} (z)$, $n>1$ the integration constants are fixed by using 
appropriate boundary conditions. For $\ell>1$, the difference equation determines $R_{\ell,n}$ in terms of $R_{\ell',n'}$ with $\ell<\ell'$. 

In the same way, the full free energy will be given by 
\be
\label{fullf}
F(t,g_s) = \sum_{\ell=0}^{\infty}  C^{\ell} F^{(l)}(t,g_s), 
\ee
where
\be
\label{fkinst}
 F^{(\ell)}(z,g_s)=  \re^{-\ell  A(t)/g_s} F_{\ell,1}(z) \Bigl( 1 +\sum_{n=1}^{\infty} g_s^{n} F_{\ell,n+1}(z)\Bigr), \qquad \ell\ge 1.
 \ee
Once (\ref{transr}) is known, one can plug it in (\ref{diffeq}) to deduce the $F^{(\ell)}(t,g_s)$. This amplitude is the $\ell$-instanton 
amplitude of the full matrix model. For example, in the case of the cubic matrix model, it gives $ \log Z^{(\ell)}(t,g_s)$, where 
$Z^{(\ell)}(t,g_s)$ is the partition function appearing in (\ref{znl}). Notice that, as in the case of differential equations, with the method sketched above one 
obtains again a one--parameter family of solutions parametrized by a constant $C$. This constant 
plays the same role as the $\theta$ parameter $\zeta$ of the original matrix model. 

The idea of looking at 
trans-series solutions of the pre-string equation to obtain 
instanton corrections in the full matrix model has not been fully exploited in the literature, but it has been appeared in related contexts.  
\cite{lattice} uses essentially 
this approach to obtain the instanton action in the strongly coupled phase of the unitary matrix model. In \cite{akk}, the instanton 
action of compactified $c=1$ string theory is obtained by considering a trans-series ansatz for a difference equation of the Toda type. 
In the beautiful paper \cite{sy}, trans-series solutions to the recursion equation for $r_n$ in the quartic matrix model 
are studied in some detail, but their focus is on the double-scaling limit.

The method based on a trans-series solutions to the difference equation has two main drawbacks as compared to other methods. 
First, it does not give the value of the Stokes parameter, which can be computed in the saddle-point method. Second, it does not 
give the most general multi-instanton expansion for the original matrix model, since it automatically incorporates 
symmetries of the potential. For example, in the case of the quartic matrix model with potential 
(\ref{quarticpot}) and $\lambda>0$ there are three 
saddle points: a minimum at $x_1=0$, and two symmetric maxima at $x_2$, $x_3=-x_2$.  The most general multi-instanton amplitude will be of the form (\ref{sumz})
\be
\label{quartictous}
Z(\zeta_2, \zeta_3)=\sum_{N_1+N_2+N_3=1} \zeta_2^{N_2} \zeta_3^{N_3} Z(N_1, N_2, N_3),
\ee
where we have fixed the overall normalization by setting $\zeta_1=1$. 
The partition function  $Z(N_1, N_2, N_3)$ describes the situation where $N_i$ 
eigenvalues sit at $x_i$. However, with the method based on orthogonal polynomials, we find only a one-parameter family depending on a single constant $C$,
\be
\label{zetaco}
Z(C)=\sum_{\ell=0}^{\infty} C^{\ell} Z^{(\ell)}.
\ee
Due to the symmetry of the problem, it is easy to see that (\ref{zetaco}) gives the partition function $Z(\zeta_2, \zeta_3)$ with $\zeta_2=\zeta_3 =C$, 
i.e. $Z^{(\ell)}$ is a symmetrized instanton amplitude
\be
Z^{(\ell)}=\sum_{N_2+N_3=\ell} Z(N-\ell, N_2, N_3).
\ee
In simple cases where symmetry is not an issue, 
like the cubic matrix model or the GWW model, the method provides however 
the full multi-instanton amplitudes in a much more efficient way than 
alternative methods. 

It is worth noting that the relation between the expansion (\ref{transr}) and (\ref{fullf}), as encoded in (\ref{diffeq}), is rather complicated. In 
order to extract explicit results for the free energy it is then useful to make it more explicit. For the $\ell$-th instanton correction, with 
$\ell \ge 1$, it follows 
from (\ref{diffeq}) that
\be
 F^{(\ell)}(z+g_s,g_s)+ F^{(\ell)}(z-g_s,g_s)-2  F^{(\ell)}(z,g_s) = \biggl[ {R^{(\ell)}(z,g_s)  \over R^{(0)} (z,g_s)}\biggr]^{c}, 
\ee
where the superscript ${\rm c}$ denotes the connected piece, i.e. 
\be
\ba
\biggl[ {R^{(\ell)}(z,g_s)  \over R^{(0)} (z,g_s)}\biggr]^{c}&= \sum_{s\ge 1} {(-1)^{s-1}\over s} \sum_{\ell_1 +\cdots + \ell_s=\ell}  
{R^{(\ell_1)}(z,g_s)  \over R^{(0)} (z,g_s)}\cdots 
 {R^{(\ell_s)}(z,g_s)  \over R^{(0)} (z,g_s)} \\
 &= {R^{(\ell)}(z,g_s)  \over R^{(0)} (z,g_s)} -{1\over 2} \sum_{k=1}^{\ell-1} {R^{(k)}(z,g_s) R^{(\ell-k)}(z,g_s)  \over (R^{(0)} (z,g_s))^2}+ \cdots. 
 \ea
 \ee
This quantity will have an expansion similar to (\ref{kpert}),
\be
\biggl[ {R^{(\ell)}(z,g_s)  \over R^{(0)} (z,g_s)}\biggr]^{c}=c_\ell(z) \Bigl( 1+\sum_{n=1}^{\infty} g_s^{n} c_{\ell,n+1}(z) \Bigr), 
\ee
and we obtain the following relations for the one- and two-loop contributions to $F^{(\ell)}(z,g_s)$: 
\be
\label{exfr}
\ba
F_{\ell,1}(z) &={1\over 4}  c_\ell(z) {\rm csch}^2\Bigl( { \ell A'(z) \over 2}\Bigr),\\
F_{\ell,2}(z)&=c_{\ell,2}(z) + {c_\ell'(z) \over c_\ell(z)} \coth \Bigl( { \ell A'(z) \over 2}\Bigr) - \ell A''(z) \biggl( {1\over 2}
   \coth^2\Bigl( \ell {A'(z) \over 2}\Bigr) +{1\over 4} {\rm csch}^2\Bigl( \ell {A'(z) \over 2}\Bigr) \biggr).
\ea
\ee
Equations for $F_{\ell,n}$, $n\ge3$ can be easily obtained from (\ref{diffeq}).

\subsection{An example: the quartic matrix model}

As a first example, we will study in some detail multi-instanton corrections in the quartic matrix model with 
potential (\ref{quarticpot}). The perturbative solution (\ref{zeroex}) has been much studied since 
it was first worked out in the pioneering papers \cite{bessis,biz}. The planar part is given by
\be
R_{0,0}(z)={2\over \lambda} \Bigl(1-{\sqrt{1-\lambda z}}\Bigr). 
\ee
As already noticed in \cite{biz}, it turns out to be useful to express all results in terms of 
\be
r=R_{0,0}(z). 
\ee
For the higher $g_s$ corrections one finds, 
\be
\ba
R_{0,2}(z)&={2 \lambda^2 \over 3} {r\over (2-\lambda r)^4},\\
R_{0,4}(z)&={28 \lambda^4\over 9} {r (5 + \lambda r) \over (2- \lambda r)^9},\\
R_{0,6}(z)&= {4 \lambda^6 \over 27} \frac{r \left(111 \lambda^2 r^2+5728 \lambda r+7700\right)}{(2-\lambda r)^{14}},
\ea
\ee
and so on. We also recall that the double-scaling limit of $R^{(0)}(z,g_s)$ is obtained at the critical value $\lambda=1$ and
\be
\label{qdsl}
g_s\rightarrow 0, \quad z\rightarrow 1, \qquad \kappa^{5\over 2}=(1-z)^{5\over 2} g_s^{-2}. 
\ee
In this limit, 
\be
u(\kappa)=g_s^{-{2\over 5}} \Bigl( 2-R^{(0)}(z,g_s)\Bigr)
\ee
satisfies the Painlev\'e I equation (\ref{p1}) as a consequence of (\ref{diff}). 

If we now plug in the trans-series ansatz (\ref{transr}) in the difference equation (\ref{diff}), we find a system of recursive difference equations  
for the $R^{(k)}(z,g_s)$:
\be
R^{(k)}(z,g_s)={\lambda\over 12} \sum_{\ell=0}^k R^{(k-\ell)}(z,g-s)\Bigl( R^{(k)}(z+g_s,g_s)+R^{(k)}(z-g_s,g_s)+R^{(k)}(z,g_s)\Bigr).
\ee
For $k=1,2$ we have, for example,
\be
\label{konetwo}
\ba
& R^{(1)}(z+g_s,g_s)+R^{(1)}(z-g_s,g_s) \\ 
&\qquad + { R^{(1)}(z,g_s) \over R^{(0)}(z,g_s)} \Bigl( 
2 R^{(0)}(z,g_s)+R^{(0)}(z+g_s,g_s)+R^{(0)}(z-g_s,g_s) -{12\over \lambda} \Bigr)
=0, \\
& R^{(2)}(z+g_s,g_s)+R^{(2)}(z-g_s,g_s)  \\
&\qquad + { R^{(2)}(z,g_s) \over R^{(0)}(z,g_s)} \Bigl( 
2 R^{(0)}(z,g_s)+R^{(0)}(z+g_s,g_s)+R^{(0)}(z-g_s,g_s) -{12\over \lambda} \Bigr) \\
&\qquad+ { R^{(1)}(z,g_s) \over R^{(0)}(z,g_s)} \Bigl( 
R^{(1)}(z+g_s,g_s)+R^{(1)}(z-g_s,g_s) \Bigr)
=0, \ea
\ee
Using now the ansatz (\ref{kpert}) we can solve for the different quantities. Let us focus on 
$k=1$, the one instanton solution. The first thing to compute is $A(z)$, which 
corresponds physically to the instanton action. From the equation for $k=1$ we find, at leading order in $g_s$, 
\be
\re^{A'(z)}+ \re^{-A'(z)} +4-{12\over \lambda r}=0, 
\ee
which gives immediately
\be
\cosh (A'(z)) = 2 {3-\lambda r \over \lambda r}. 
\ee
This can be integrated to find $A(z)$ up to an additive constant and an overall sign (since $\cosh\, z$ is even). Both 
ambiguities can be fixed by requiring that, near the critical point, 
\be
A(z) \sim {4 {\sqrt 6} \over 5} (1-z)^{5\over 4}.
\ee
The result is
 \be
 \ba
A(z)&= - \int  \rd r \cosh^{-1} \Bigl( 2 {3-\lambda r \over \lambda r}\Bigr) \Bigl(1-{\lambda r\over 2}\Bigr)\\
= &\frac{1}{4} r  (\lambda r-4) \cosh ^{-1}\left(\frac{6}{\lambda r}-2\right)+{1\over 2 \lambda}
   \sqrt{3 (2- \lambda r)(6-\lambda r)}.
   \ea
   \label{quarticinstanton}
 \ee
It can be checked that (\ref{quarticinstanton}) coincides with the instanton action 
of the quartic matrix model computed in terms of its spectral curve in \cite{msw}. Notice that $z$ stands here for the 
't Hooft parameter. 

Once the instanton action is known, we can proceed to compute $R_{1,1}(z)$. The equation one obtains at the next order in $g_s$ is
\be
{R'_{1,1}(z) \over R_{1,1}(z)}=-{1\over 2} \coth (A'(z)) A''(z), 
\ee
which can be immediately integrated as
\be
R_{1,1}(z)= \Bigl( \sinh (A'(z))\Bigr)^{-1/ 2}. 
\ee
The rest of the coefficients can be found by integrating the resulting equations for $R_{1,n}(z)$, and one 
finds for example, up to three loops, 
\be
\ba
R_{1,2}(z)&=-\frac{\lambda^3 r^3-6 \lambda^2 r^2+6 \lambda r+24}{2 \sqrt{3} r (2-\lambda r)^{5/2} (6-\lambda r)^{3/2}},\\
R_{1,3}(z)&=\frac{17 \lambda^6 r^6-268 \lambda^5 r^5+1800 \lambda^4 r^4-5688 \lambda^3 r^3+6660 \lambda^2 r^2+288 \lambda r+576}{24 r^2 (\lambda r-6)^3 (\lambda r-2)^5}.
\ea
\ee
This result can be checked by using the double-scaling limit (\ref{qdsl}), since with the above values
\be
1+ \sum_{n=1}^{\infty} g_s^n R_{1,n+1}(z) \rightarrow u^{(1)}(\kappa)=1 -{5\over 32 {\sqrt {6}}} \kappa^{-{5\over 4}} + {75\over 4096} \kappa^{-{5\over 2}} -\cdots, 
\ee
which are indeed the first terms of $u^{(1)}$, the one-instanton trans-series solution to Painlev\'e I (\ref{p1}) (in order to compare to eq. 4.40 in \cite{msw}, one has to rescale $g_s \rightarrow g_s/{\sqrt{2}}$). 

Using the results for $R_{1,n}$, $n\ge 1$, 
as well as (\ref{exfr}) we find for the one-instanton, one-loop free energy, 
\be
\label{qoneloop}
F^{(1)}_{1,1}(z)={1\over 2 r} \Bigl( \cosh (A'(z)) -1\Bigr)^{-{5\over 4}} \Bigl( \cosh (A'(z)) +1\Bigr)^{-{1\over 4}} =  {\lambda ^ {3\over 2} r^{1\over 2} \over 2 \Bigl(3(2- \lambda r)\Bigr)^{5/4} (6-\lambda r)^{1\over 4}},
\ee
while for the two and three-loop contributions we have
\be
\label{qtwothree}
\ba
F_{1,2}(z)&=-\frac{5 \lambda^3 r^3-54 \lambda^2 r^2+150 \lambda r+24}{2 \sqrt{3} r (2-\lambda r)^{5/2} (6-\lambda r)^{3/2}},\\
F_{1,3}(z)&=\frac{25 \lambda^6 r^6-828 \lambda^5 r^5+10008 \lambda^4 r^4-50424 \lambda^3 r^3+89028 \lambda^2 r^2+7200 \lambda r+576}{24 r^2 (\lambda r-6)^3 (\lambda r-2)^5}.
\ea
\ee
We can now compare these results to those obtained in \cite{msw}. The one-loop calculation 
in \cite{msw} computes $F^{(1)}_{1,1}(z)$, times a coefficient. This coefficient is precisely 
the Stokes parameter $S$ that gives the 
discontinuity as we cross a Stokes line. Comparing (\ref{qoneloop}) with the result in \cite{msw}, we 
find complete agreement with the functional dependence on $z$, and we also find that in order to match 
the Stokes parameter we have to set 
\be
S={\sqrt{ 3 g_s \over 2\pi \lambda}}. 
\ee
The result at two loops in (\ref{qtwothree}) fully agrees with the one presented in \cite{msw}. 

\sectiono{Nonperturbative effects in the unitary matrix model}

\subsection{$1/N$ expansion and phase structure}
We will first review some well--known aspects of the $1/N$ expansion of unitary matrix 
models. For a detailed account with many references see for example \cite{rv}. 

As in the Hermitian case, the planar limit of a unitary matrix model is described by a density for the eigenvalues of the unitary matrix, 
$\rho(\phi)$, which verifies the normalization condition
\be
\label{norm}
\int_{-\pi}^{\pi} \rho(\phi) \rd\phi=1. 
\ee
The free energy of the unitary matrix model has a $g_s$ expansion of the 
standard form (\ref{largeNf}), and the planar free energy $F_0$ can be computed in terms of the density as
\be
\label{uplanarf}
\ba
F& =t \int_{-\pi}^{\pi} \rd\phi \, \rho(\phi) V(\phi) + {t^2 \over 2} \int_{-\pi}^{\pi} \rd\phi  \int_{-\pi}^{\pi} \rd\psi \, \rho(\phi) \rho(\psi) 
\log \biggl[ 4 \sin^2  \Bigl( {\phi- \psi \over 2} \Bigr)\biggr]  \\ & + \xi\Bigl( \int_{-\pi}^{\pi} \rd\phi\,  \rho(\phi) -1 \Bigr) 
\ea
\ee
where $t=N g_s$ is the 't Hooft parameter and $\xi$ is a Lagrange multiplier which imposes the constraint (\ref{norm}). 
The density $\rho$ satisfies the equation 
\be
\label{usaddle}
{1\over t} V(\phi) + \int_{-\pi}^{\pi} \rd\psi  \, \rho(\psi) 
\log \sin^2  \biggl( {\phi- \psi \over 2} \biggr) + \xi =0,
\ee
which implies that the effective potential
\be
V_{\rm eff}(\phi)=-V(\phi) - t \int_{-\pi}^{\pi} \rd \psi\, \rho(\psi)  \log \biggl[ \sin^2  \Bigl( {\phi- \psi \over 2} \Bigr) \biggr],
\ee
is constant on the support of $\rho$. 

As first found in \cite{gw}, unitary matrix models have a rich phase structure (see \cite{jz,mandal} for a detailed discussion). In the so-called 
{\it ungapped} phase, the density of eigenvalues has its support on the entire circle and is of the form \cite{mandal,jz}
\be
\rho(\phi)={1\over 2 \pi} \biggl( 1 + \sum_l \Bigl( l g_l \re^{\ri l \phi} + {\rm c.c.} \Bigr) \biggr).
\ee
In the {\it one-gap} phase, the support of the density of eigenvalues is a single, connected interval inside $[-\pi, \pi]$, and the density 
of eigenvalues is of the form
\be
\label{gaprho}
\rho(\alpha) ={1\over 2\pi}  g(\alpha) \cos {\alpha\over 2} {\sqrt{\sin^2{\alpha_c\over 2} -\sin^2 {\alpha\over2}}}, 
\ee
and is supported on the interval 
\be
\CC = [-\alpha_c, \alpha_c] \subset [-\pi, \pi].
\ee
The effective potential can be computed as
\be
\label{uveffalt}
V_{\rm eff}(\phi)-V_{\rm eff}(\alpha_c) =t \int_{\alpha_c}^{\phi} \rd z \, g(z) \cos {z\over 2} {\sqrt{\sin^2{z\over 2} -\sin^2 {\alpha_c\over2}}}.
\ee

A simple model which exhibits two phases is the famous GWW model \cite{gw,wadia}. The potential is simply
\be
V(z)= {1\over 2}(z + z^{-1}).
\ee
In the ungapped phase, the density of eigenvalues is
\be
\label{strongdens}
\rho (\phi)= {1\over 2 \pi} \biggl(1 + {1\over t} \cos \phi \biggr).
\ee
It is easy to see that this density is positive 
as long as $t>1$, which is strong 't Hooft coupling. Therefore we will refer to the 
the ungapped phase as the {\it strongly coupled} phase of the model. 
For $t=1$, (\ref{strongdens}) vanishes at $\phi=\pi$, and for $t<1$ it becomes 
negative around $\phi=\pi$ and it is no longer acceptable as a solution. There must be a phase transition at $t=1$, and 
for $t <1$, i.e. in the {\it weakly coupled} phase, the density of eigenvalues takes the form (\ref{gaprho}) with $g(z) =2/t $:
\be
\label{gwrho}
\rho (\phi)={1\over \pi t} \cos \Bigl({\phi\over 2}\Bigr) {\sqrt {t - \sin^2{\phi\over 2}}}, \qquad t<1, 
\ee
for $\phi \in [-\alpha_c, \alpha_c]$, while it vanishes outside this interval. The endpoint of the support is determined by the condition 
\be
\sin^2{\alpha_c\over 2}=t.
\ee
In \figref{gwwtrans} we show the form of the density $\rho(\phi)$ as we go through the transition at $t=1$. 

\FIGURE[!ht]{
\leavevmode
\centering
\epsfysize=4cm
 \epsfbox{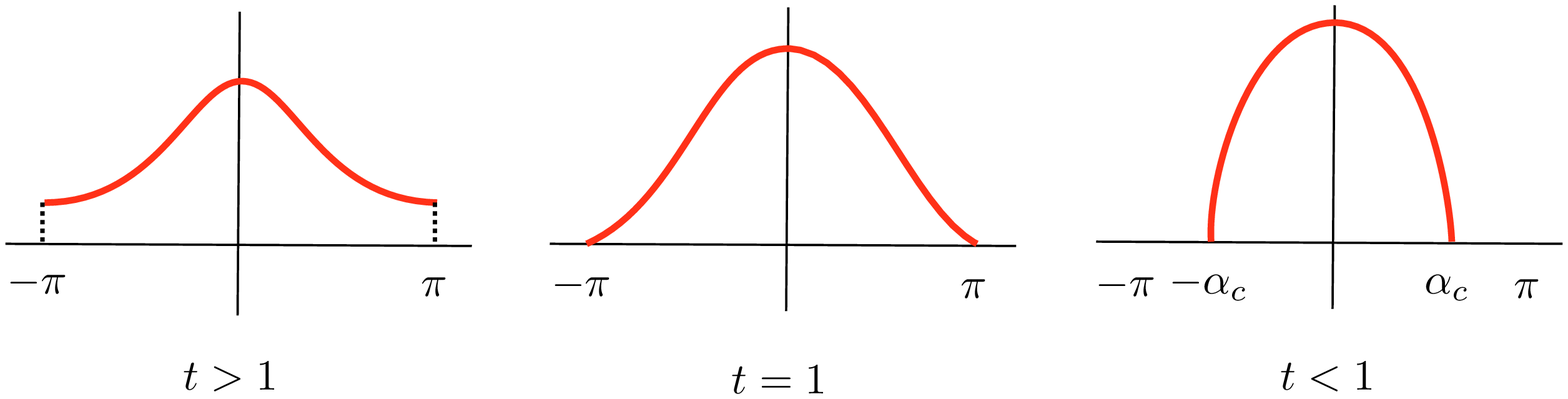}  \hspace{3cm}
\caption{The density of eigenvalues $\rho(\phi)$ in the strongly coupled phase (left), at the transition point (center) and in the 
weakly coupled phase (right).}
\label{gwwtrans}
}

Using the above densities we can easily calculate the planar free energies in both phases, and one finds
\be
\ba
F^{\rm w}_0(t) &= {t^2 \over 2} \Bigl( \log\, t -{3\over 2}\Bigr) + t, \quad & t<1, \\
F^{\rm s}_0(t)&={1\over 4}, \quad & t>1.
\ea
\ee
Since the free energy and its two first derivatives are continuous at $t=1$, we have a third order phase transition at large $N$ \cite{gw,wadia}. 

The GWW model can be studied as well by using the method of orthogonal polynomials \cite{lattice}. One first derives a recursion relation 
for the coefficients $f_n$ appearing in (\ref{fnrec}), which reads
\be
\label{gwwfrec}
g_s(n+1) f_n ={1\over 2} (1-f_n^2) (f_{n+1} +f_{n-1}). 
\ee
In the continuum limit 
\be
\label{fcont}
f_n \rightarrow f(z, g_s) 
\ee
and the recursion (\ref{gwwfrec}) becomes the difference equation
\be
\label{udiff}
(z+g_s) f(z,g_s)={1\over 2} (1-f^2(z,g_s))(f(z+g_s,g_s)+ f(z-g_s,g_s)).
\ee
In order to compute the partition function we need the continuum limit of the coefficients $r_n$, $R(z,g_s)$, 
which is related to $f(z,g_s)$ by the continuum counterpart of (\ref{hf}), 
\be
\label{rfrel}
R(z,g_s) =1-f^2(z-g_s,g_s).
\ee
One then deduces the following difference equation for $R(z,g_s)$, 
\be
\label{diffR}
z {\sqrt {1-R(z,g_s)}} ={1 \over 2} R(z,g_s) \biggl\{ {\sqrt{1-R(z + g_s,g_s)}} + {\sqrt{1-R(z-g_s,g_s)}}\biggr\}.
\ee
This difference equation has two different solutions depending on the value of $z$, which reflect the existence of two phases in the model \cite{lattice}:
\be
\label{rphase}
 R^{(0)}(z,g_s)=\begin{cases} \sum_{\ell=0}^{\infty} R_{0,2\ell}(z) g_s^{2\ell} &\text{if $z<1$}\\
 1 &\text{if $z\ge1$}.
 \end{cases}
 \ee
 For the solution in the region $z<1$ one finds
 \be
 \ba
R_{0,0}(z) &=z, \\
R_{0,2}(z)&={1\over 8} {z\over (1-z)^2}, \\
R_{0,4}(z)&=\frac{9 z (z+3)}{128 (1-z)^5},
\ea
\ee
and so on. The reason for the existence of these two solutions can be easily understood if one looks at (\ref{hf}). This equation implies that
$0 \le r_n\le 1$. At leading order in $g_s$, the solution $R_{0,0}(z)=z$ already violates this constraint, therefore when $z=1$ 
the bound is saturated and we must have $R_{0,0}(z)=1$. One then finds that there are no $g_s$ corrections to this solution, and 
one ends up with (\ref{rphase}). 

The method of orthogonal polynomials gives as well a very efficient way of computing the perturbative free energies in the weakly coupled phase. In \cite{lattice} 
results up to genus $2$ were obtained, but going to higher genus is just a matter of CPU time. It is convenient to compute the normalized free energies, where 
one subtracts the free energies of the Gaussian matrix model. In this way one obtains, 
\be
\ba
F_0(t)&=t, \\
F_1(t)&=-{1\over 8} \log (1-t), \\
F_2(t)&={3 t \over 128 (1-t)^3}, \\
F_3(t)&={9 t (5+2t) \over 1024 (1-t)^6}, \\
\ea
\ee
and so on. This results are valid for $t<1$. 

It is usually stated in the literature that the free energy in the strong coupling phase is given by its planar part, plus nonperturbative corrections 
coming from instantons. In fact, {\it both} phases have instanton corrections, albeit of a different character, and 
we will study both of them by considering trans-series solutions to the difference equation (\ref{diffR}).

\subsection{Instanton corrections in the GWW model}

We now use the method explained in section 3 to compute the instanton corrections in both phases of the GWW model. 

\FIGURE[!ht]{
\leavevmode
\centering
\epsfysize=4cm
\hspace{3cm}\epsfbox{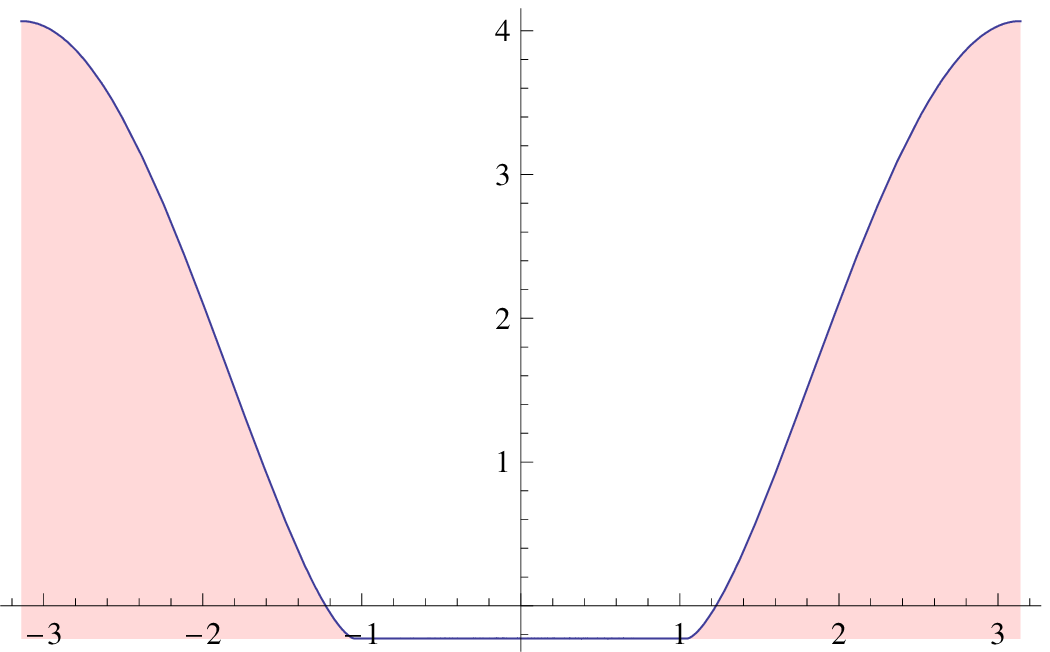}\hspace{3cm}
\caption{The effective potential for the GWW model, for 't Hooft parameter $t=1/4$. The support of the eigenvalue distribution, where 
the effective potential is constant, is the interval $[-\pi/3, \pi/3]$.}
\label{effpot}
}

We first consider the weakly coupled phase. In this phase instanton corrections have an easy interpretation in terms of 
eigenvalue tunneling. The potential for the eigenvalues of the unitary matrix is $-V(\theta)=-\cos\theta$, which has a minimum at $\theta=0$ and a 
maximum at $\theta =\pi \equiv -\pi$. The effective potential taking into account the eigenvalue repulsion can be computed as
\be
V_{\rm eff}(\theta)-V_{\rm eff}(\alpha_c)  = 4 t  \Phi \biggl( {\sin {\theta\over 2}\over \sin {\alpha_c\over 2}},
\biggr)
\ee
where
\be
\Phi(x) ={1\over 2} x {\sqrt{x^2-1}}- {1\over 2} \cosh^{-1}(x).
\ee
This effective potential is of the form shown in \figref{effpot} and its maximum is still at $\theta=\pi$. Therefore, there will be multi-instanton 
configurations obtained by taking $\ell$ eigenvalues from the support of the density of eigenvalues $[-\alpha_c, \alpha_c]$, centered around the 
minimum at $\theta=0$, to $\theta=\pi\equiv-\pi$, as shown in \figref{weakinst}. 
The action of such an instanton can be easily computed, as a function of the 't Hooft parameter, to be
\be
\label{ainstoff}
A(t)=V_{\rm eff}(\pi)-V_{\rm eff}(\alpha_c) =4 t  \Phi \bigl( t^{-{1\over 2}} \bigr).
\ee

\FIGURE[!ht]{
\leavevmode
\centering
\epsfysize=6cm
\hspace{3cm}\epsfbox{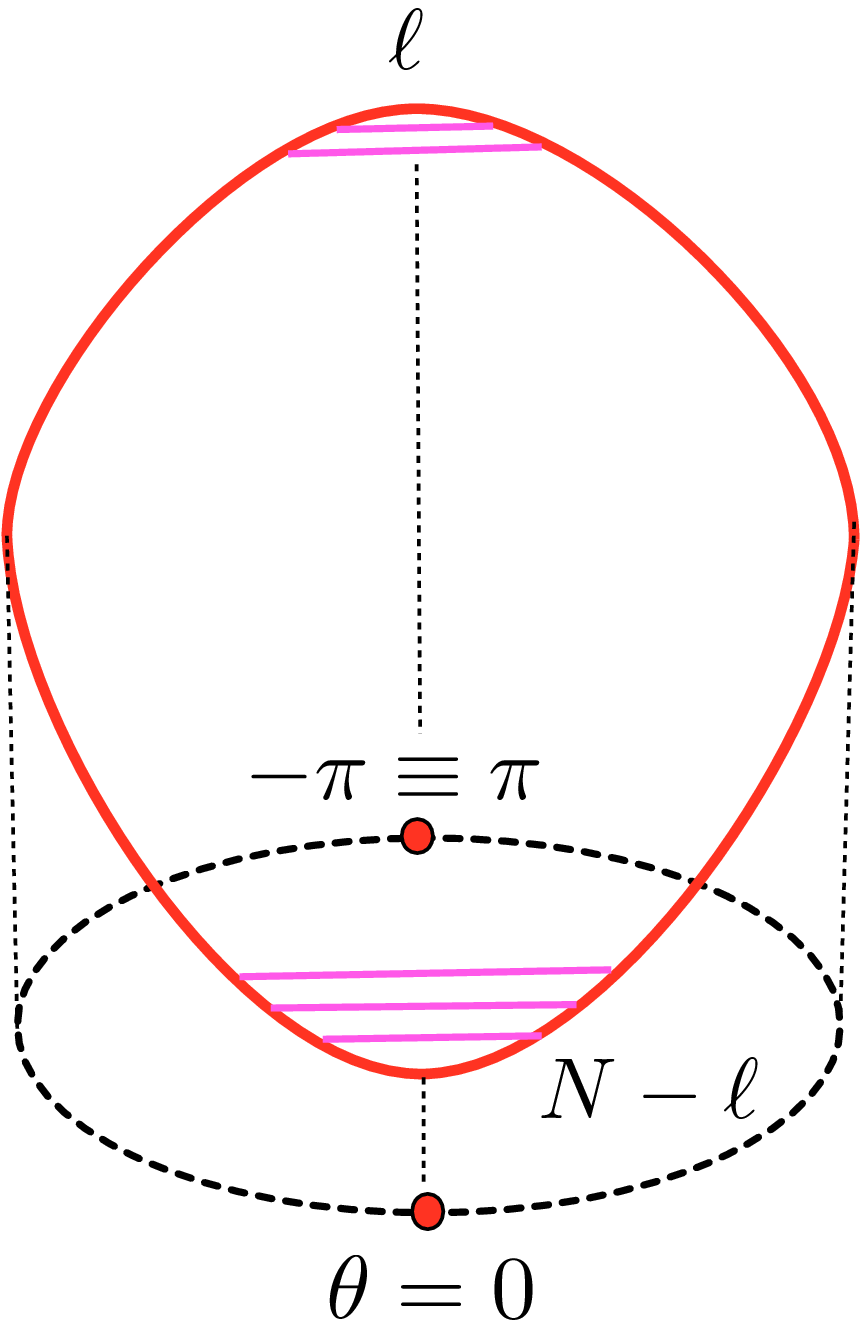}\hspace{3cm}
\caption{An $\ell$-instanton configuration in the weakly coupled phase of the GWW model. The eigenvalues of the unitary matrix live on a circle parametrized by an angle $\theta \in [-\pi, \pi]$, and the potential is $-\cos \theta$. There are $N-\ell$ eigenvalues sitting around the minimum at $\theta=0$, while $\ell$ eigenvalues sit at the maximum at $\theta=\pi \equiv-\pi$.}
\label{weakinst}
}

In order to find the perturbative expansion around these instantons one can use the approach followed in \cite{msw}. In fact, it is possible to map the unitary matrix model into a Hermitian matrix model \cite{mizoguchi} 
where the calculations of \cite{msw} could in principle be used 
{\it verbatim}. However, when one does this, the saddle point at $\theta=\pi$ is sent to infinity and this leads to extra subtleties in the 
evaluation of the saddle-point integral. On top of that, going beyond the one-instanton, two-loop calculation of \cite{msw} seems 
rather hard. 

Instead of following this strategy, we will obtain a trans-series solution to the difference equation (\ref{diffR}), which makes 
possible to calculate any instanton amplitude. By plugging the trans-series ansatz (\ref{transr}) in (\ref{diffR}) we obtain, as before, a 
series of recursive relations for the $R^{(k)}(z,g_s)$. For $k=1$ we obtain the difference equation, 
\be
\label{linun}
\ba
& R^{(1)} (z,g_s) \Bigl\{ {z\over {\sqrt{1-R^{(0)}(z,g_s)}}} + {\sqrt{1-R^{(0)}(z+g_s,g_s)}} + 
{\sqrt{1-R^{(0)}(z-g_s,g_s)}} \Bigr\}\\
&-{1 \over 2} R^{(0)}(z,g_s) \Bigl\{ {R^{(1)}(z+g_s,g_s) \over {\sqrt{1-R^{(0)}(z+g_s,g_s)}}}+ 
{R^{(1)}(z-g_s,g_s) \over  {\sqrt{1-R^{(0)}(z-g_s,g_s)}}} \Bigr\} =0.
\ea
\ee
This can be easily solved for $R^{(1)}(z,g_s)$, to all orders in $g_s$. First, we solve for $A(z)$, which is determined by the leading 
order of (\ref{linun}) as a series in $g_s$. We find
\be
\cosh (A'(z)) = {2\over z}-1, 
\ee
which gives
\be
\label{uninstz}
A(z)=-z \cosh^{-1} \Bigl( 
2/z -1\Bigr) +2 {\sqrt{1-z}}, \qquad z<1.
\ee
As in the quartic matrix model, we have fixed the ambiguities by requiring the right behavior at the critical point $z=1$.
This agrees with the result (\ref{ainstoff}) obtained with the effective potential for the eigenvalues. 
At order $\CO(g_s)$ we get, 
\be
{R_{1,1}'(z) \over R_{1,1}(z)} =-{1\over 2} {1\over 1-z} -{A''(z) \over 2} \coth(A'(z)), 
\ee
which can be immediately integrated to 
\be
R_{1,1}(z) =  (1-z)^{1\over 2} \sinh^{-{1\over 2}} (A'(z)),
\ee
up to a multiplicative integration constant that can be reabsorbed in $C$. Since
\be
\sinh(A'(z))=- {\sqrt{ (2/z-1)^2 -1}} = - {2\over z} {\sqrt {1-z}}.
\ee
we finally obtain 
\be 
R_{1,1}(z)= z^{1\over 2} (1-z)^{1\over 4}. 
\ee
As we explained above, we have in fact a one-parameter family of solutions parametrized by a constant 
$C$ as in (\ref{transr}). We will determine the corresponding value of the Stokes multiplier in the next subsection, by 
comparing the instanton amplitude to known results in the double-scaling limit. 

The functions $R_{1,n}$, corresponding to $n$-loop amplitudes around the instanton 
solution, can be now computed in a straightforward way to any order $n$. We present 
results for $n=2,3,4$:
\be
\ba
R_{1,2}(z)&=\frac{3 z^2-12 z-8}{96 (1-z)^{3/2} z},\\
R_{1,3}(z)&=\frac{81 z^4-2376 z^3+2400 z^2+192 z+64}{18432 (1-z)^3 z^2},\\
R_{1,4}(z)&=\frac{30375 z^6-208980 z^5+281880 z^4-4078080 z^3+289728 z^2-343296 z+71168}{26542080 (1-z)^{9/2} z^3},
\ea
\ee
which are valid in the weakly coupled phase with $z<1$. We can then use this result to compute the one-instanton contribution to the free energy, 
by using (\ref{diffeq}). One finds, up to three loops, 
\be
\label{ufreeone}
\ba
F^{(1)}(z,g_s)&={1\over 4} z^{1\over 2} (1-z)^{-3/4}\biggl[ 1 +\frac{3 z^2-60 z-8}{96 (1-z)^{3/2} z} g_s \\
\qquad &+\frac{81 z^4+792 z^3+17376 z^2+960 z+64}{18432 (1-z)^3 z^2} 
g_s^2 +\cdots \biggr] \re^{-A(z)/g_s}, \qquad z<1,
\ea
\ee
where $z$ stands here of course for the 't Hooft parameter and $A(z)$ is given by (\ref{uninstz}). 

We now consider the strong coupling phase of the model for $z>1$. This is an ungapped phase where 
the eigenvalues fill the circle, therefore we can no longer interpret instanton effects in terms of eigenvalue 
tunneling. We can however find a trans-series solution to the difference equations in this phase (and in fact, some 
ingredients of this method were already sketched in \cite{lattice}). Since we are 
expanding around a different perturbative solution, the difference equations for the 
instanton amplitudes change. We obtain, for $k=1$, 
\be
4 z^2 R^{(1)} (z,g_s) =R^{(1)}(z+g_s,g_s)+ R^{(1)}(z-g_s,g_s)+ 2 {\sqrt{ R^{(1)}(z+g_s,g_s)R^{(1)}(z-g_s,g_s)}}.
\ee
Again, we solve first for the equation determining the instanton action, which in this 
case is simply
\be
\cosh\Bigl({A'(z)\over 2}\Bigr)=z, 
\ee
and leads to 
\be
\label{stronginst}
A(z) =2 z \cosh^{-1}(z) -2 {\sqrt{z^2-1}}, \qquad z>1,
\ee
in agreement with the result of \cite{lattice}. It is also easy to compute $R^{(1)}(z,g_s)$ 
to any order. We write the result up to four loops, 
\be
\ba
R^{(1)}(z,g_s)&= (z^2-1)^{-1/2} \biggl[ 1-\frac{2 z^2+3}{12 \left(z^2-1\right)^{3/2}} g_s +\frac{4 z^4+156 z^2+45}{288 \left(z^2-1\right)^3}g_s^2 \\
 \qquad & +\frac{248 z^6-31716 z^4-73602 z^2-8505}{51840 \left(z^2-1\right)^{9/2}}g_s^3 +\cdots\biggr],
 \ea
 \ee
Of course, by using the above result it is straightforward to compute the free energy. We also 
note that, although we have given explicit results for the one-instanton amplitude only, it is straightforward to calculate higher instanton 
corrections. 

\FIGURE[!ht]{
\leavevmode
\centering
\epsfysize=4cm
\hspace{3cm}  \epsfbox{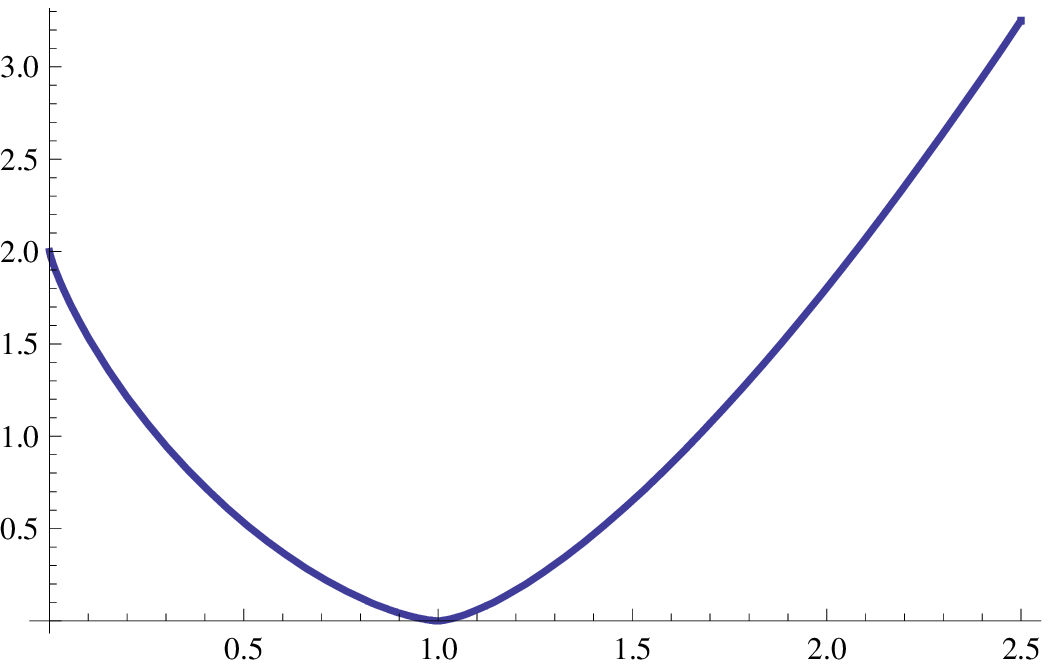} \hspace{3cm}
\caption{The instanton action in the GWW model, as a function of the 't Hooft parameter $t$. For $0<t<1$ it is given by (\ref{uninstz}), and for $t>1$, after the 
phase transition, it is given by (\ref{stronginst}). The action vanishes at the critical point $t=1$, and it is real and positive for all $t>0$.}
\label{iaction}
}

One important aspect of our result is that the instanton action {\it vanishes} at the critical point $z=1$ (see \figref{iaction}). We can then say that 
the third order phase transition discovered in \cite{gw,wadia} is triggered by instantons, as first conjectured in \cite{neuberger}, since 
instanton corrections become of order $1$ at the transition point. This instanton-driven transition has been found before in two models closely 
related to the GWW model. The first one is the quantum-mechanical unitary model introduced and solved by Wadia in \cite{qmwadia} (this is the main example 
studied by Neuberger in \cite{neuberger}). The second one is continuum 
two-dimensional Yang--Mills theory on the sphere \cite{gma}. In fact, it is interesting to observe that the action of 
the instanton in the weakly coupled phase of 2d Yang--Mills theory is 
identical to (\ref{uninstz}) after the identification 
$z=A/\pi^2 $, where $A$ is the area of the sphere and $z$ is the 't Hooft coupling in the GWW model (we note that in order to compare with \cite{gm}, where the exponentiated instanton 
action is written like $\exp(-NA)$, one has to divide (\ref{uninstz}) by $z$, since in our conventions we write it like $\exp(-A/g_s)$). 

 \subsection{Multi-instantons and the double-scaling limit}

We now discuss formal trans-series solutions in the double scaling limit of the GWW model, which will provide (among 
other things) a check of the above results.
 
The double-scaling limit of the GWW model is defined by  
\be
\label{ukappa}
g_s \rightarrow 0, \quad t \rightarrow 1, \qquad \kappa= g_s^{-{2\over 3}} (1-t) \, \, {\rm fixed}, 
\ee
and describes the universal scaling near the third-order phase transition. 
In this limit, the function 
\be
u(\kappa)=g_s^{-{1\over 3}} f(t,g_s),
\ee
where $f(t,g_s)$ is defined by (\ref{fcont}), satisfies the Painlev\'e II equation 
\be
\label{p2}
 u''(\kappa)-2 u^3(\kappa)+ 2\kappa u(\kappa)=0
\ee
as a consequence of the difference equation (\ref{udiff}). The double--scaled free energy 
$F_{\rm ds}(\kappa)$ is defined as the double-scaling limit of
\be
F^{\rm s}(t,g_s) -F^{\rm w}(t,g_s) 
\ee
and satisfies
\be
\label{uds}
F''_{\rm ds}(\kappa) =u^2(\kappa). 
\ee
Notice that the regions $\kappa \rightarrow \pm \infty$ are mapped to $t\rightarrow 1^{\mp}$, 
therefore they correspond to the weak and the strong coupling phase, respectively. 

We first discuss the double-scaling limit of the weakly coupled phase. Since
\be
f(z,g_s) = {\sqrt {1-z}} +\CO(g_s), 
\ee
it follows that the solution to Painlev\'e II which describes the double-scaling limit of the unitary 
matrix model must behave 
like
\be
\label{plusas}
u(\kappa) \sim {\sqrt{\kappa}}, \quad \kappa \rightarrow \infty.
\ee
This asymptotic behavior determines a unique formal solution to (\ref{p2}) of the form  
\be
\label{pertp2}
u^{(0)}(\kappa)= {\sqrt{\kappa}} - 
  \frac{1}{16\,\kappa^{\frac{5}{2}}} - \frac{73}{512\,\kappa^{\frac{11}{2}}}- 
  \frac{10657}{8192\,\kappa^{\frac{17}{2}}}  - \frac{13912277}{542888\,\kappa^{\frac{23}{2}}} 
+\cdots, \qquad \kappa \rightarrow \infty.
\ee
As in the case of Painlev\'e I, one can consider as well exponentially suppressed corrections to this perturbative behavior and construct a formal trans-series solution with the structure,
\be
\label{p2trans}
u(\kappa) =\sum_{\ell=0}^{\infty} C^{\ell} u^{(\ell)}(\kappa)={\sqrt {\kappa}}  
\sum_{\ell=0}^{\infty} C^{\ell} \kappa^{-{3 \ell \over 4}} \re^{-\ell A \kappa^{3/2}} \epsilon^{(\ell)}(\kappa), \quad \kappa \rightarrow \infty,
\ee
where
\be
A={4\over 3}
\ee
and
\be
\label{eltwo}
\epsilon^{(\ell)}(\kappa)=\sum_{n=0}^{\infty} u_{\ell,n+1} \kappa^{-3n/2}.
\ee
As before, we normalize the solution with $u_{1,1}=1$. 
The perturbative part $u^{(0}(\kappa)$ is given by (\ref{pertp2}). The instanton expansions can be easily found by plugging the trans-series 
ansatz in the Painlev\'e II equation. One finds a recursive equation of the form, 
\be
(u^{(n)})''+2 \kappa u^{(n)}-2 \sum_{k_1+k_2+k_3=n} u^{(k_1)} u^{(k_2)} u^{(k_3)}=0. 
\ee
For example, $u^{(1)}$ satisfies the linear equation, 
\be
 (u^{(1)})'' + 2 \kappa u^{(1)} -6 (u^{(0)})^2 u^{(1)}=0, 
 \ee
 while $u^{(2)}$ satisfies the equation
 \be
 (u^{(2)})'' + 2 \kappa u^{(2)} -6 (u^{(0)})^2 u^{(2)}=6 u^{(0)} (u^{(1)})^2,
 \ee
 and their asymptotic expansion as $\kappa \rightarrow \infty$ are given by 
 \be
 \ba
\epsilon^{(1)}(\kappa)&=1-{17\over 96} \kappa^{-3/2} + {1513 \over 18432} \kappa^{-3} -\cdots,\\
\epsilon^{(2)}(\kappa)&={1\over 2}-{41\over 96} \kappa^{-3/2} + {5461\over 9216} \kappa^{-3} -\cdots.
\ea
\ee

(\ref{p2trans}) gives a one-parameter family of formal solutions to the Painlev\'e II equation which includes  
exponentially small corrections. Since the positive real axis for $\kappa$ is a Stokes line for this problem, 
we can ask what is the value of the Stokes parameter. This can be obtained by 
various methods. One option is a direct one-loop computation in a matrix model whose critical behavior is described by Painlev\'e II. Such a 
computation has been done in \cite{kawai} for the symmetric, quartic matrix model with two colliding cuts. Alternatively, one can use results in the 
theory of isomonodromy deformations \cite{kapaev,painlevet}. One finds
\be
S =-{\ri \over {\sqrt{2\pi}}}. 
\ee

We can now use the results about the trans-series solution of Painlev\'e II for $\kappa \rightarrow \infty$ to test some of the results that 
we obained for the unitary matrix model in the weakly coupled region. As a 
consequence of (\ref{rfrel}), we have in this region, 
\be
\label{compar}
R^{(1)}(z,g_s) = -2 f^{(0)}(z,g_s) f^{(1)}(z,g_s), 
\ee
where $f^{(\ell)}(z,g_s)$ is the $\ell$-instanton contribution to the full $f(z,g_s)$. 
Therefore, in the double-scaling limit we should have that
\be
R^{(1)}(z,g_s) \rightarrow -2 u^{(0)}(\kappa) u^{(1)}(\kappa). 
\ee
Indeed, one verifies from the explicit results presented above that the instanton action (\ref{uninstz}) behaves like, 
\be
A(z) \sim {4\over 3} (1-z)^{3\over 2}, \quad z\rightarrow 1^-
\ee
and that
\be
R_{1,1}(z) \sim (1-z)^{1\over 4}, \quad z\rightarrow 1^-, 
\ee
in agreement with (\ref{compar}). Moreover, we can fix in this way the Stokes multiplier for $R(z,g_s)$ in the weakly coupled phase
\be
\label{weaks}
S^{\rm w} =\ri {\sqrt {2 g_s\over \pi}}. 
\ee
Finally, one can check that 
\be
1+ \sum_{n=1}^{\infty} g_s^n R_{1,n+1}(z) \rightarrow \epsilon^{(0)}(\kappa)\epsilon^{(1)}(\kappa)=1-{17\over 96} \kappa^{-3/2} + {361 \over 18432} 
\kappa^{-3} -{791441 \over 5308416} \kappa^{-9/2} +\cdots.
\ee

Let us now discuss the double scaling limit of the strongly coupled phase. In this phase, due to (\ref{rphase}) and (\ref{rfrel}) we have 
that $f^{(0)}(z,g_s)=0$ and 
\be
f^{(1)}(z,g_s)={\sqrt {R^{(1)}(z,g_s)}}.
\ee
The instanton action (\ref{stronginst}) behaves like,
\be
A(z) \sim {4 {\sqrt{2}} \over 3} (z-1)^{3\over 2}, \quad z\rightarrow 1^+,
\ee
therefore the relevant solution to Painlev\'e II must behave as,
\be
\label{minusas}
u(\kappa) \sim \re^{-\tilde A(-\kappa)^{3/2}}, \quad \kappa \rightarrow -\infty,
\ee
where 
\be
\tilde A={2 {\sqrt 2} \over 3}.
\ee
As before, we can consider exponentially small corrections to the asymptotics and construct 
a one-parameter family of formal trans-series solutions with the structure, 
\be
\label{strongtrans}
u(\kappa) =\sum_{\ell=1}^{\infty} D^{\ell} \tilde u^{(\ell)}(\kappa) ={\sqrt {\kappa}}  \sum_{\ell=1}^{\infty} 
D^{\ell} (-\kappa)^{-{3 (2 \ell+1) \over 4}} \re^{-(2 \ell +1)\tilde A (-\kappa)^{3/2}} \tilde \epsilon^{(\ell)}(\kappa),\quad \kappa \rightarrow -\infty,
\ee
where 
\be
\label{elm}
\tilde \epsilon^{(\ell)}(\kappa)=\sum_{n=0}^{\infty} \tilde u_{\ell,n+1} (-\kappa)^{-3n/2}
\ee
and we normalize $\tilde u_{1,1}=1$. It follows that $\tilde u^{(1)}(\kappa)$ satisfies, 
\be
(\tilde u^{(1)}(\kappa))''+2 \kappa \, \tilde u^{(1)}(\kappa)=0,
\ee
which is, up to normalization, the Airy equation. Using the well-known asymptotics for the function ${\rm Ai}(z)$ (see for example \cite{bo}, pp. 101--102) we find, 
\be
\tilde \epsilon^{(1)}(\kappa)= 1+\sum_{n=1}^{\infty}\Bigl( -{3\over 4 {\sqrt{2}}}\Bigr)^n  {\Gamma\Bigl( n+{1\over 6}\Bigr) \Gamma\Bigl( n+{5\over 6}\Bigr) 
\over n! \Gamma\Bigl({1\over 6}\Bigr) \Gamma\Bigl({5\over 6}\Bigr)} (-\kappa)^{-3n/2}, \quad \kappa \rightarrow -\infty.
\ee
Using this we can check the result for $R^{(1)}(z,g_s)$ in the full unitary model the strongly coupled region, since
\be
1+ \sum_{n=1}^{\infty} g_s^n R_{1,n+1}(z) \rightarrow (\tilde\epsilon^{(1)})^2(\kappa)=1 -\frac{5 }{24 \sqrt{2}}(-\kappa)^{-3/2} + \frac{205 }{2304}
(-\kappa)^{-3} -\frac{22715}{165888 \sqrt{2}}(-\kappa)^{-9/2}+\cdots.
\ee
So far we have only considered formal solutions to the Painlev\'e II equation, and we have obtained two one-parameter 
families of trans-series solutions characterized by their asymptotic behavior, 
namely (\ref{p2trans}) and (\ref{strongtrans}). It turns out \cite{hm} that there is a unique, actual solution 
to (\ref{p2}) which belongs to both families and has the asymptotic behaviors (\ref{plusas}), (\ref{minusas}). This is known as the {\it Hastings--McLeod solution} to Painlev\'e II, and it 
defines the double-scaling limit nonperturbatively \cite{cdm}. In the next section we will discuss the 
Hastings --McLeod solution and its relation to 
the formal trans-series, as well as the extension of this structure to the unitary matrix model off-criticality.

\subsection{Large order behavior}
An important application of instanton calculus is the determination of the large order behavior 
of perturbation theory. As in \cite{msw}, we can now use the results on the one-instanton 
correction in the weakly coupled phase of the unitary matrix model to determine the 
large order behavior of the $1/N$ series. More precisely, knowledge of the one-instanton 
contribution $F^{(1)}(z,g_s)$ and of the Stokes multiplier determines a $1/g$ asymptotic 
expansion for the genus $g$ free energies $F_g(z)$. The precise formula is 
\be
\label{lostring}
\ba
F_g (z)  \sim &{A(z)^{-2g-b} \over \pi}\, \Gamma(2g+b) (-\ri S^{\rm w}) F_{1,1}(z)  \\
& \quad \cdot \left[1 + {A(z) F_{1,2}(z) \over 2g+b-1} + {A^2(z) F_{1,3}(z) \over (2g+b-2)(2g+b-1)} + \cdots \right],
\ea
\ee
where $b=-5/2$. 

This large order formula can be tested numerically 
by analyzing the sequence of $F_g(z)$ for sufficiently large $g$ and by removing tails with the 
use of Richardson transforms. The test goes as follows \cite{msw}. We first use the finite 
sequence 
\be
F_g (z), \quad g=0, 1, \cdots N
\ee
 to construct
\be\label{richainst}
Q_g(z)={F_{g+1}(z)\over 4g^2F_g(z)}, \quad g=0, \cdots, N-1.
\ee
If (\ref{lostring}) holds, this sequence should have the asymptotic behavior
\be
Q_g(z) \sim {1\over A^2(z)}\left(1+{1+2b\over 2g}+\CO\left({1\over g^2} \right)\right)
\ee
as $g\rightarrow \infty$. We can now use Richardson transforms to extract the value of $A$ from the $k$-th Richardson transform of the sequence $\{ Q_g \}_{g=0, \cdots, N-1}$, which we denote by 
\be
Q_g^{(k)}(z), \qquad g=0,1, \cdots, N-k-1. 
\ee
The best estimate of $A(z)$ with this sequence is then 
\be
A^{(k)}(z)={1\over {\sqrt{Q^{(k)}_{N-k-1}(z)}}}. 
\ee
The functions $F_{1,n}(z)$ are extracted in a similar way. For example, for $F_{1,1}(z)$ and $F_{1,2}(z)$ we consider the sequences given by 
\be\label{rich2}
\ba
S_{1,g}(z)&= {\pi A(z)^{2g+b}F_g(z) \over (-\ri S^{\rm w}) \Gamma(2g+b)} \rightarrow F_{1,1}(z), \quad g\rightarrow \infty\\
S_{2,g}(z)&={2g\over A(z)}\left({\pi A(z)^{2g+b}F_g(z) \over (-\ri S^{\rm w}) F_{1,1}(z) \Gamma(2g+b)}-1 \right)
\rightarrow F_{1,2}(z), \quad g\rightarrow \infty,
\ea
\ee
as well as their Richardson transforms $S_{i,g}^{(k)}(z)$. 
This produces the numerical estimates
\be
F^{(k)}_{1,i}(z) =S_{i,N-k-1}^{(k)}(z), \quad i=1,2
\ee
for the one-loop and two-loop functions in terms of the $1/N$ expansion. 

\FIGURE[ht]{
    \centering
    \epsfxsize=0.5\textwidth
    \leavevmode
    \mbox{\hspace{-1cm}\epsfbox{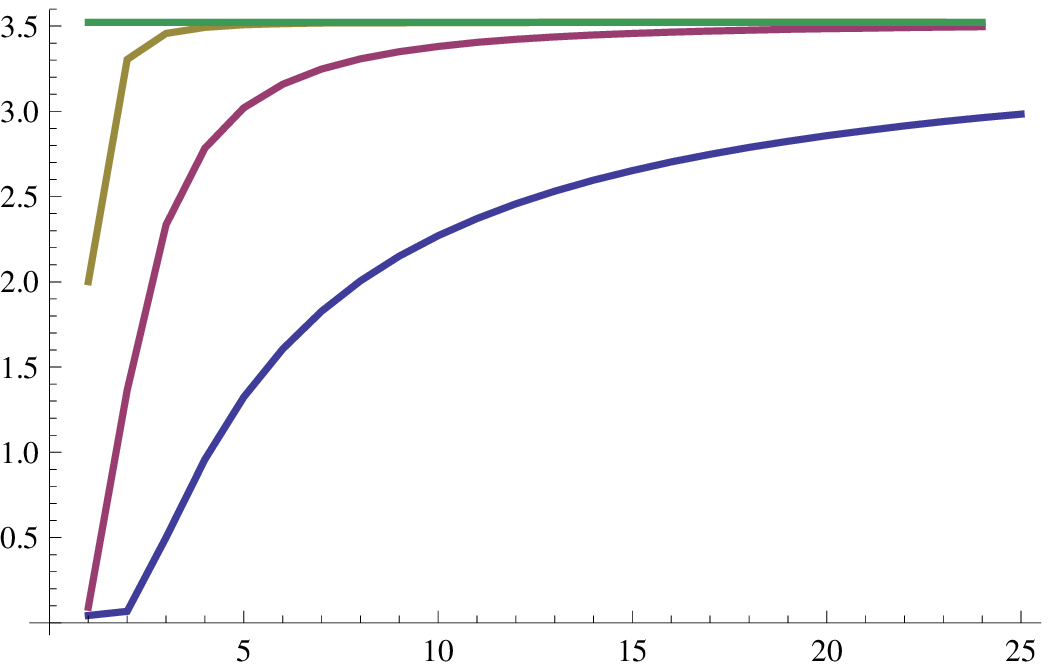}\quad
    \epsfbox{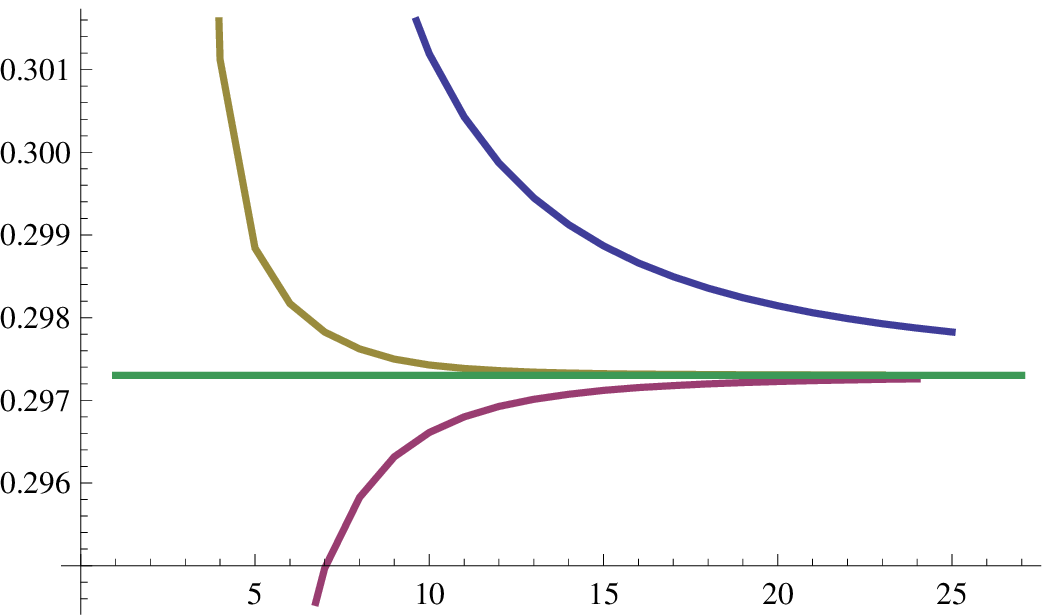}\hspace{-1cm}}
\caption{The figure in the left shows the sequence $Q_g$ defined in (\ref{richainst}), as well as its Richardson transforms $Q^{(k)}_g$, $k=1,2$, evaluated for $z=1/2$ and with $N=26$. The horizontal line at the top is the value of the inverse squared instanton action $1/A^2(1/2)$, where $A(z)$ is given by (\ref{uninstz}). The figure on the right shows the sequence $S_{1,g}$ as defined in (\ref{rich2}) and its Richardson transforms $S^{(k)}_{1,g}$, again for $z=1/2$ and $N=26$. The horizontal line is the value of $F_{1,1}(1/2)$, where 
$F_{1,1}(z)=z^{1/2} (1-z)^{-{3/4}}$ is the one loop prefactor in (\ref{ufreeone}).}
\label{fig:qainst}
}

In \figref{fig:qainst} and \figref{fig:twoloop} we compare the sequences $Q_g$, $S_{i,g}$, $i=1,2$ and their Richardson transforms for $k=1,2$, to the analytic results 
for the inverse squared instanton action, $F_{1,1}(z)$ and $F_{1,2}(z)$, respectively, for $z=1/2$. The analytic results are plotted as vertical lines. Already 
from the plots we see that the agreement 
between the analytic prediction and the actual large order behavior is extremely good. Just to give some numerical examples, the estimate for the instanton action coming from the fifth Richardson transform, evaluated at $z=1/2$, is
\be
A^{(5)}(1/2)=0.5328399880
\ee
while the exact result is
\be
A(1/2)=0.5328399754. 
\ee
Similarly, we have, for the one and two-loop estimates as compared to the exact result, 
\be
\ba
F^{(5)}_{1,1}(1/2)=&0.2973018513, \qquad F^{(5)}_{1,2}(1/2)=&-2.194973650,\\
F_{1,1}(1/2)=&0.2973017788, \qquad F_{1,2}(1/2)=&-2.194977300.
\ea
\ee

\FIGURE[ht]{
\leavevmode
 \centering
 \epsfxsize=0.5\textwidth
  \hspace{3cm} \epsfbox{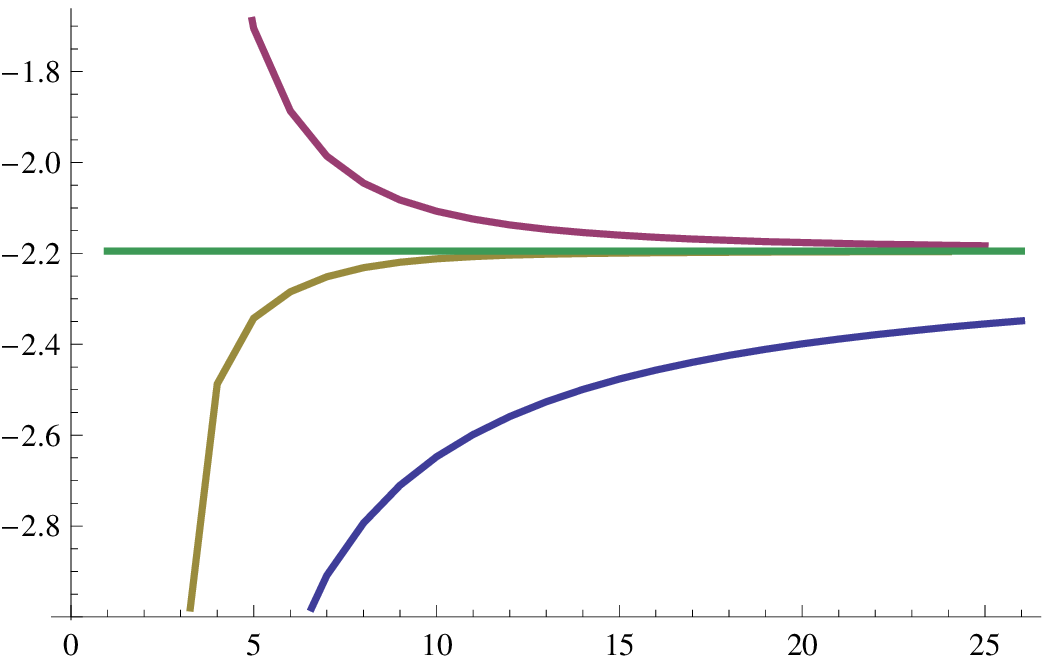} \hspace{3cm}
\caption{The sequence $S_{2,g}$ as defined in (\ref{rich2}) and its Richardson transforms $S^{(k)}_{2,g}$, $k=1,2$, again for $z=1/2$ and $N=26$. The horizontal line is the value of $F_{1,2}(1/2)$.}
\label{fig:twoloop}
}

This analysis confirms, indeed, that the free energies $F_g$ of the GWW model in the weakly coupled phase diverge factorially. It is easy to verify that the 
instanton action (\ref{uninstz}) which controls the large order behavior is real and positive in the full phase $0<t<1$ (see \figref{iaction}). Therefore, the $1/N$ expansion 
of the free energy $F$ in the weakly coupled phase is, technically speaking, {\it not} Borel summable, since its Borel transform 
will have singularities at points of the form $\ell A$, $\ell=1, 2, \cdots$. We will discuss this issue in much more detail in the next section.

\section{Nonperturbative effects and nonperturbative definitions}

The genus expansion and its instanton corrections give only {\it formal}, asymptotic expansions of the original matrix integrals. In fact, as power series 
in $g_s$, these series are badly divergent. One can see that 
\be
F_g(z) \sim (2g)!, \qquad F_{\ell,n}(z) \sim n!, \quad \ell\ge 1. 
\ee
On the other hand, the original integrals are in many cases well--defined. In the unitary case this is clear, since the 
integration over unitary matrices is over a compact domain. In the Hermitian case, the general partition function 
(\ref{sumz}) is a linear combination of convergent integrals. The question we will address in this section is how 
to recover the original, convergent matrix integrals, from the formal trans-series solutions. We will focus on the unitary case, 
since it has been comparatively less discussed, but we will start by analyzing the problem in the double-scaling limit, where 
we can rely on known results in the theory of exponential asymptotics and of resurgent functions.

\subsection{Formal solutions and Borel resummation}

As so often in quantum mechanics and quantum field theory, in order to give a meaning to the formal series we have obtained we must 
resum them. There are many resummation techniques available in the literature (see \cite{useful} for a recent review), but since 
the series we have to deal with diverge factorially, it is natural to use Borel resummation. We now briefly review 
some basic ideas of Borel resummation and of the theory of resurgent functions. 

Let 
\be
\phi(w) =\sum_{n=0}^{\infty} a_n w^n 
\ee
be a factorially divergent series, where
\be
a_n \sim (\beta n)!. 
\ee
The {\it Borel transform} of $\phi$, $\widehat \phi(z)$, is defined as the series 
\be
\widehat \phi(z)=\sum_{n=0}^{\infty} {a_n \over (\beta n)!} z^n.
\ee
This series defines typically a function which is analytic in a neighboorhood of the origin. If (1) the resulting function can be analytically continued to a neigbourhood 
of the positive real axis, and (2) the integral
\be
\label{borelsum}
f(w)= w^{-1/\beta} \int_0^{\infty} \rd t \, \re^{-t/w^{1/\beta}} \widehat \phi (t^{\beta})
\ee
converges in some region of the $w$-plane, the series $\phi(w)$ is said to be {\it Borel summable} in that 
region. In that case, $f(w)$ defines a function whose asymptotics 
coincides with the original, divergent series $\phi(w)$, and $f(w)$ is called the {\it Borel sum} of $\phi(w)$. 

In many cases of physical interest (like in quantum field theory and in the examples considered 
in this paper), one finds that $\widehat \phi(z)$ can be analytically continued but it develops 
singularities (poles or branch cuts) along the real axis. This is also a typical situation in the analysis of 
irregular singular points of differential equations. 
Traditionally, the appearance of singularities on the real axis is regarded as an obstruction to Borel summability, since the integral 
(\ref{borelsum}) is ill-defined and a prescription has to be given in order to avoid the singularities. But one can still use 
Borel resummation in order to construct well-defined quantities, and indeed this is the main problem addressed in the theory 
of resurgent functions and in the theory of exponential asymptotics. In the following we will rely very much on results coming from these theories. 
The theory of resurgent functions has been developed by \'Ecalle in his monumental work \cite{ecalle}, but for our purposes 
the results presented in \cite{approche,costin,seara,delapham} will be enough.

\FIGURE[!ht]{
\leavevmode
\centering
\epsfysize=4cm
  \hspace{3cm}\epsfbox{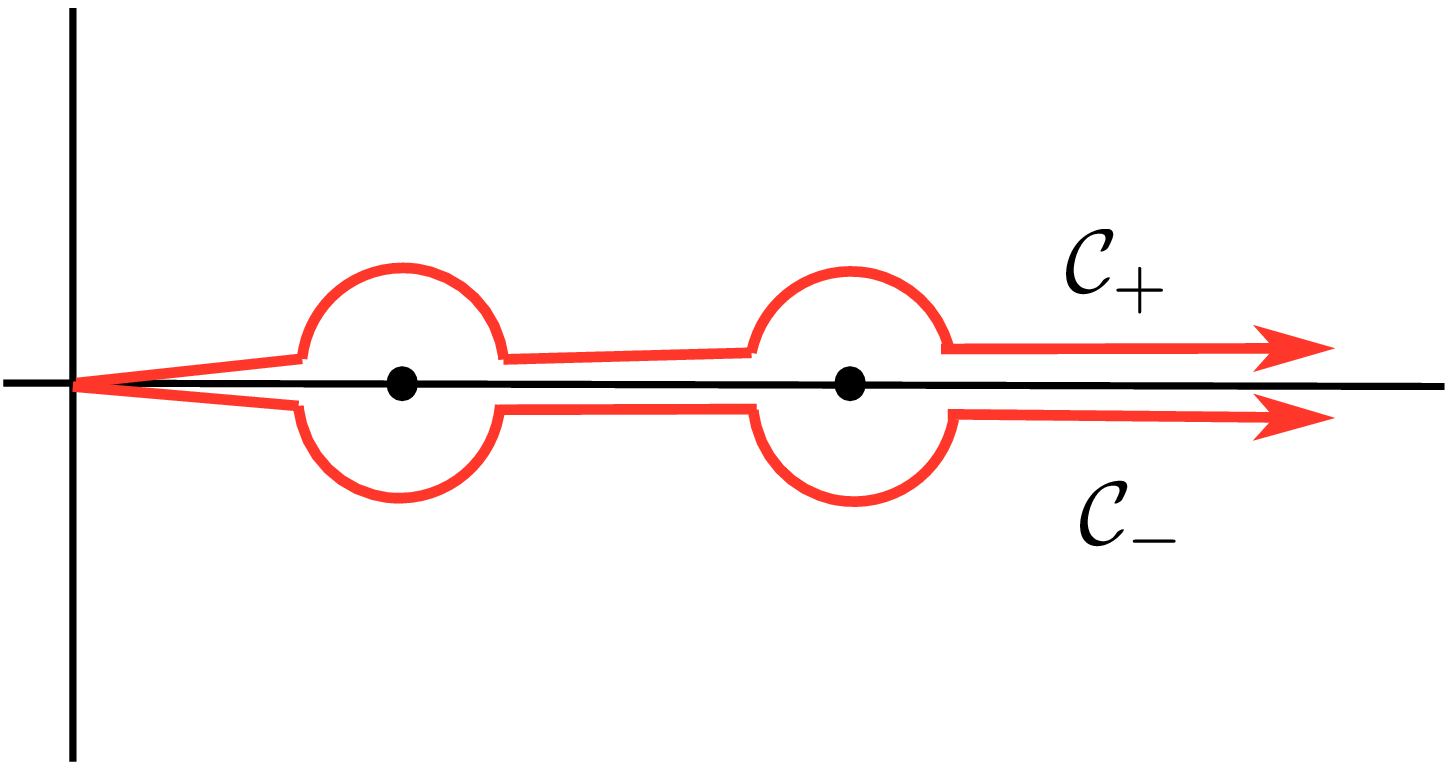}  \hspace{3cm}
\caption{The paths ${\cal C}_{\pm}$ avoiding the singularities of the Borel transform from above (respectively, below).}
\label{lateralfig}
}

In order to avoid the singularities, we will consider {\it lateral Borel resummations}. Let ${\cal C}_{\pm}$ be a path going from $0$ to $\infty$ and avoiding 
the singularities of $\widehat \phi(z)$ on the real axis from above (resp. below). Typically, these paths have the form shown in \figref{lateralfig}. 
The lateral Borel resummations are then defined as
\be
(s_{\pm} \phi)(w)= w^{-1/\beta} \int_{{\cal C}_{\pm}} \rd t \, \re^{-t/w^{1/\beta}} \widehat \phi (t^{\beta}),
\ee
provided the integral is convergent. Notice that, even if the original series has real coefficients, since the lateral Borel resummations are computed by integrals 
along paths in the complex plane, they lead in general to complex-valued functions. The resummations from above and from below are related 
by complex conjugation
\be
\label{conj}
\CH s_+ =s_-\CH, 
\ee
where $\CH$ is the Hermitian conjugation operator $(\CH f)(z)=\bar f (\bar z)$. Lateral resummations play a central role in the theory of resurgent functions, and they 
have been also used to resum nonalternating perturbative series in a variety of problems in quantum mechanics and quantum field theory \cite{fgs,jentschura}. 
In particular, it has been shown that in many cases the imaginary part which is obtained 
when doing these resummations has a physical interpretation. For example, 
the perturbation series for the ground state energy of the cubic oscillator is not Borel summable, 
since the Borel transform exhibits a singularity in the 
positive real axis. In this case, the lack of Borel summability is just reflecting the instability of the potential. 
The lateral resummations lead in this case to a complex answer for the Borel sum of the energy, but this is as it should be, since the energy of the ground state should have an imaginary 
part which gives the width of the level (see for example \cite{alvarez, useful}).

We are interested in the divergent series which appear in the context of matrix models, namely when solving a differential or difference equation. Let us first focus on 
the case of differential equations. We will have in mind the cases of interest for us, namely the Painlev\'e I and II differential equations. The results which we will use 
were obtained in \cite{ecalle,costin, approche} and they hold for a large class of nonlinear differential equations with an irregular singular point 
at infinity (see \cite{costin} for a precise statement of the theorems and their conditions of validity). 

As we have seen in the case of Painlev\'e I and II, along the  
Stokes line ${\rm arg}(\kappa)=0$ one can construct a family of formal trans-series solutions of the form
\be
\label{formalt}
u(\kappa;C) =\sum_{\ell=0}^{\infty} C^{\ell} u^{(\ell)}(\kappa), 
\ee
where
\be
u^{(\ell)}(\kappa) = \kappa^{\alpha + \ell \beta } \re^{-\ell A \kappa^{\gamma}} \epsilon^{(\ell)}(\kappa), 
\ee
$\alpha, \beta, \gamma$ are characteristic exponents of the differential equation, $A >0$ is a constant, and 
\be
\epsilon^{(\ell)}(\kappa) =\sum_{n=0}^{\infty} u_{\ell,n+1} \kappa^{-\gamma n}
\ee
are asymptotic series diverging like $u_{\ell,n+1}\sim n!$. The Borel transforms of these series can be defined as above, 
by identifying $z=\kappa^{-\gamma}$, and they have singularities at the 
points in the positive real axis of the form $\ell A$, with $\ell=1, 2, \cdots$. We can avoid these singularities by performing 
lateral Borel resummations of all the formal power series appearing in the formal solution. In this way we construct the functions
\be
\label{latsols}
u_{\pm}(\kappa;C_{\pm})=\sum_{\ell=0}^{\infty} C_{\pm}^{\ell} u^{(\ell)}_{\pm}(\kappa) = 
\sum_{\ell=0}^{\infty} C_{\pm}^{\ell} \kappa^{\alpha + \ell \beta } \re^{-\ell A \kappa^{\gamma}} (s_{\pm}\epsilon^{(\ell)})(\kappa).
\ee

It turns out that, if ${\rm Re}\, \kappa>0$ is sufficiently big, the infinite sum over $\ell$ is convergent in some angular sector around the 
real axis. Therefore, the lateral Borel resummations produce {\it true}, no longer formal, solutions of the original differential equation 
in this sector. Moreover, 
any solution of the differential equation with the asymptotics given by the formal series $u^{(0)}(\kappa)$ as $\kappa \rightarrow \infty$ 
can be represented in the form (\ref{latsols}) for some $C_{\pm}$ \cite{costin}. This is an important point, since we could think of 
many ways of avoiding the singularities of the Borel transform along the real axis, by choosing different contours. However, the 
general results on this type of equations tells us that the use of lateral resummations in the way we have explained is already enough 
to generate all the relevant solutions. In fact, we have already too many solutions, since 
the two contours $\CC_{\pm}$ in \figref{lateralfig} produce {\it two} 
different families with the same asymptotic behavior, therefore they should be related. The relation 
is given by (see \cite{approche, costin})
\be
\label{resurgence}
u_+(\kappa;C) =u_-(\kappa;C+S), 
\ee
where $S$ is the Stokes parameter (which is purely imaginary). This equation 
gives an infinite number of relations between the functions $u^{(\ell)}_{\pm}$ which can be obtained by 
taking derivatives on both sides w.r.t. $C$ and then setting $C=0$. In this way one finds, 
\be
u_+^{(\ell)} -u^{(\ell)}_-=\sum_{k=1}^{\infty} {\ell+k \choose \ell} S^k u_-^{(\ell+k)}.
\ee
This expresses the fact that the lateral Borel resummations of the $\ell$-instanton correction 
differ by an imaginary part which is exponentially suppressed with respect to their real parts. Moreover, at leading 
order this imaginary part is proportional to the $(\ell+1)$-instanton correction. For $\ell=0,1$ we obtain, for example, 
\be
\label{resdet}
\ba
u_+^{(0)} -u^{(0)}_-&=S u^{(1)}_-+\cdots ={S\over 2}(u^{(1)}_++u^{(1)}_-) + \cdots, \\
u_+^{(1)} -u^{(1)}_-&=2 S u^{(2)}_-+\cdots =S(u^{(2)}_++u^{(2)}_-) + \cdots,\\
\ea
\ee
where the dots denote higher order instanton contributions. 

It is clear that the one-parameter families of solutions (\ref{latsols}) are in general not real, even if the 
starting point were divergent series with real coefficients. However, the resurgence relation (\ref{resurgence}) makes possible 
to construct a one-parameter family of solutions which are manifestly real for $\kappa \in \IR$. This family is given by
\be
\label{realsol}
u^{\IR}(\kappa; C) =u_+(\kappa;C-S/2), \qquad C\in \IR. 
\ee
To see this, notice that (\ref{resurgence}) gives, for $C \rightarrow C-S/2$, 
\be
\label{corresign}
u_+(\kappa;C-S/2) =u_-(\kappa;C+S/2).
\ee
therefore, if we use (\ref{conj}) we obtain
\be
\Bigl( u^{\IR}(\kappa; C)\Bigr)^*=u_-(\kappa;C+S/2)=u^{\IR}(\kappa; C), \quad \kappa \in \IR,
\ee
so (\ref{realsol}) is real. In \'Ecalle's theory, the solution (\ref{realsol}) is called the {\it median resummation} of the 
formal trans-series (\ref{formalt}). Of course, in this solution 
all imaginary parts coming from Borel resummation cancel in the end. We have
\be
\label{expand}
u^{\IR}(\kappa; C) =u_+^{(0)}(\kappa) + (C-S/2) u_+^{(1)}(\kappa)+\cdots
\ee
so at first order in the exponential factor $\re^{-A \kappa^{\gamma}}$ the imaginary part of (\ref{expand}) is given by
\be
\label{cancellation}
{\rm Im}\, u_+^{(0)}(\kappa)+  \ri {S \over 2} {\rm Re}\, u^{(1)}_+
\ee
which cancels due to the first relation in (\ref{resdet}). Higher order imaginary terms also cancel, and 
using these cancellations we can write the expansion of (\ref{realsol}), up to three instantons, in 
a form where the reality properties are manifest, 
\be
\label{manifreal}
\ba
 u^{\IR}(\kappa; C) &={1\over 2} (u^{(0)}_+ +u^{(0)}_-) + {C \over 2} (u^{(1)}_+ +u^{(1)}_-) + {1\over 2} \Bigl( C^2 -{S^2\over 4}\Bigr) (u^{(2)}_+ +u^{(2)}_-) \\
 &+{1\over 2}C\Bigl( C^2 -{3 S^2\over 4}\Bigr) (u^{(3)}_+ +u^{(3)}_-)  +\cdots.
 \ea
 \ee

The cancellation taking place 
in (\ref{cancellation}) is in fact a particularly clean example of the so-called cancellation of nonperturbative ambiguities. As we have seen, 
after Borel resummation, $u^{(0)}$ picks an imaginary part which is 
ambiguous and depends on the choice of contour. In the case of the lateral resummations we have considered, this is a sign ambiguity. This ambiguity is accompanied by a similar ambiguity for the coefficient of the one-instanton contribution $u^{(1)}$, which is $C \pm S/2$. 
The relation (\ref{corresign}) tells us how these two ambiguities should be correlated in such a way 
that they cancel in the final, real answer (\ref{realsol}). 

The cancellation of nonperturbative ambiguities has been much 
discussed for renormalons \cite{davidren,grunberg} as well as for 
instantons in quantum mechanics (see \cite{zjj} for the most updated study and references to earlier work). 
In the quantum-mechanical double well \cite{zjj}, the standard perturbative series of the 
energies is not Borel summable, yet the lack of Borel 
summability is not a manifestation of an instability in the system. Therefore, there must be explicit nonperturbative 
contributions which cancel the imaginary parts incurred in when performing the Borel 
resummation. The ambiguous nonperturbative effect occurs in this case at the two-instanton level, and the final 
resummed answer is real. In the context of renormalon physics, the role of $u(\kappa)$ is played by a QCD observable 
(typically a current-current correlator where one can apply the ITEP sum rules), $u^{(0)}(\kappa)$ corresponds to 
the perturbative series, and $u^{(\ell)}(\kappa)$ correspond to nonperturbative contributions due to condensates. Consistency of the QCD path integral 
requires that the imaginary part of the Borel-resummed perturbative series cancel against the 
imaginary part of the first nontrivial condensate. In fact, this fixes the ambiguous imaginary part of the first condensate  
once a prescription is chosen for resumming the perturbative series. 

In our case, the cancellation of nonperturbative ambiguities which occurs for the median resummation (\ref{realsol}) 
is a consequence of the resurgence relation (\ref{resurgence}) and guarantees that the final answer will be real. 
However, it does not determine the constant $C$. To fix $C$ one needs further nonperturbative input. 
We will now see how the value of 
$C$ can be fixed in the unitary matrix model and its double-scaling limit. 

\subsection{The case of Painlev\'e II}

We will now apply some of the above results to the case of Painlev\'e II. From the trans-series solution (\ref{p2trans}) we can find, 
using Borel resummation, two 
one-parameter families of solutions $u_{\pm} (\kappa;C)$ to the differential equation (\ref{p2}), and all the members of these families have the right asymptotics (\ref{plusas}) as $\kappa \rightarrow \infty$. The general theory sketched above tells us that these two families are related through the resurgence relation (\ref{resurgence}), 
which leads to the cancellation of nonperturbative ambiguities. 

It is instructive to verify explicitly the relations (\ref{resdet}) by using resummation techniques. In order to do that, we compute the asymptotic expansions $\epsilon^{(\ell)}(\kappa)$, up to a given order $n$, and we form the Borel transformed series 
\be
\ba
\widehat \epsilon^{(0)}(z)&= \sum_{g=0}^{\infty} {u_{0,g} \over (2g)!} z^{2g},\\
\widehat \epsilon^{(\ell)}(z)&=\sum_{n=0}^{\infty} {u_{\ell,n+1} \over n!} z^n, \quad \ell\ge 1.
\ea
\ee
As we mentioned before, one needs to perform an analytic continuation of this series in a region including the positive real axis in order to be able to compute the 
Laplace-Borel transform (\ref{borelsum}). One way to do this in practice is to use Pad\'e approximants. We recall that, given a series 
\be
S(z)=\sum_{k=0}^{\infty} a_k z^k
\ee
the Pad\'e approximant $[l/m]$ is given by a rational function 
\be
\label{lmpade}
[l/m]_{S}(z) = {p_0 + p_1 z +\cdots + p_l z^l \over 
q_0 + q_1 z +\cdots + q_m z^m}, 
\ee
where $q_0$ is fixed to $1$, and one requires that
\be
S(z) -[l/m]_{S}(z) =\CO(z^{l+m+1}). 
\ee
This fixes the coefficients involved in (\ref{lmpade}). Given a series $\phi(z)$ we can construct the Pad\'e approximant of its Borel transform 
\be
\label{paden}
\CP^{\phi}_n(z)= \Bigl[ [n/2]/[(n+1)/2] \Bigr]_{\widehat \phi}
\ee
where $[\cdot]$ denotes the integer part (this is the approximant proposed in \cite{jentschura,useful}). $\CP^{\phi}_n(z)$ is a rational function with 
various poles on the complex plane. If the Borel transform has for example a branch cut, the Pad\'e approximant will mimick this by a series of poles along the 
cut. The first pole of the approximant will be close to the branch point of the Borel transform, and increasingly so as $n$ grows. 
A good approximation to the Borel resummed series will then be an integral of the form (\ref{borelsum}) where one 
integrates instead $\CP^{\phi}_n(z)$. In our case, we compute 
\be
\ba
 (s_{\pm} \epsilon^{(0)}_n)(w)&={1\over w^{1\over 2}} \int_{\cal C_{\pm}} \rd t \, \re^{-t/w^{1\over 2}}\CP^{\epsilon^{(0)}}_n(t),\\
 (s_{\pm} \epsilon^{(\ell)}_n)(w)&={1\over w} \int_{\cal C_{\pm}} \rd t \, \re^{-t/w}\CP^{\epsilon^{(\ell)}}_n(t), \quad \ell\ge 1,
 \ea
 \ee
where we take for $\cal C_{\pm}$ a path from $0$ to $\infty$ along the directions $\pm \pi/4$ (like for example in \cite{fgs}). 
By contour deformation, these paths should give the 
same result as the paths shown in \figref{lateralfig}, at least for the true analytic continuation of the Borel transform. 
The Pad\'e approximant can have spurious poles 
away from the real axis, and in some situations one might want to correct for these (see \cite{jentschura}). In our case, 
however, all the poles of the 
Pad\'e approximants are on the real axis, so the above integral should give a good approximation to the true result if $n$ is sufficiently large. 

\TABLE[ht]{
\centering 
\begin{tabular}{|c || c | c|} 
\hline 
$\kappa$ & $ {\rm Im} \, u_+^{(0)}(\kappa)$ & $-\ri S  \,{\rm Re} \, u_+^{(1)}(\kappa)/2 $  \\ [0.5ex] 
\hline 
$1$ & $-0.0457932$ & $-0.0457633$ \\ 
$2 $&$ -0.0036383676$ &$ -0.0036383581$ \\
$3 $& $-0.000143729160176$ & $ -0.000143729159748$ \\
$4$ & $-3.2181810964596 \cdot 10^{-6} $ & $-3.2181810964557 \cdot 10^{-6} $ \\
5 & $-4.409270574264102  \cdot 10^{-8}$ &  $-4.409270574264109 \cdot 10^{-8}$ \\ [1ex] 
\hline 
\end{tabular}
\caption{Cancellation of nonperturbative ambiguities at the one-instanton level for various values of the double-scaled 
parameter $\kappa$. At leading order in the trans-series expansion, ${\rm Im} \, u_+^{(0)}(\kappa)$ 
should be equal to $-\ri S  \,{\rm Re} \, u_+^{(1)}(\kappa)/2 $, as this table shows. The difference between both quantities is a three-instanton effect.} 
\label{table:cancone} 
}

In table \ref{table:cancone} we compare numerically ${\rm Re} \, u_+^{(0)}(\kappa)$ to $-\ri S  \,{\rm Im} \, u_+^{(1)}(\kappa)/2 $. All the computations 
have been done with $n=100$ terms in the Borel transform, which then is used to compute the Pad\'e approximant. Since we keep $n$ fixed, our approximation 
will be worst as $\kappa$ gets small, but the results in the tables are exact at the level of precision that we display. We see that both results are quite close but 
do not quite agree. This is not surprising, since they are equal only up to three-instanton corrections which we have not calculated (these are the corrections to the 
first line in (\ref{resdet})). 
These corrections go like $\exp(-4 \kappa^{3/2})$ and they become less important as $\kappa$ grows, as shown in the table. In table \ref{table:cantwo} we compare numerically ${\rm Re} \, u_+^{(1)}(\kappa)$ to $-\ri S  \,{\rm Im} \, u_+^{(1)}(\kappa) $, again for $n=100$, with similar results.

\TABLE[ht]{ 
\centering 
\hspace{3cm}
\begin{tabular}{|c || c | c|} 
\hline 
$\kappa$ & $ {\rm Im} \, u_+^{(1)}(\kappa)$ & $-\ri S  \,{\rm Re} \, u_+^{(2)}(\kappa) $  \\ [0.5ex] 
\hline 
$1$ & $-0.008163$ & $-0.008152$ \\ 
$2 $&$ -0.00004143955$ &$ -0.00004143932$ \\
$3 $& $-5.53260679 \cdot 10^{-8} $ & $ -5.53260675 \cdot 10^{-8}$ \\
$4$ & $-2.459781001 \cdot 10^{-11} $ & $ -2.459781082\cdot 10^{-11} $ \\
$5$ & $-4.1875843088  \cdot 10^{-15}$ &  $-4.1875852452 \cdot 10^{-15}$ \\ [1ex] 
\hline 
\end{tabular}
\hspace{3cm}
\caption{Cancellation of nonperturbative ambiguities at the two-instanton level.}
\label{table:cantwo}
}

We know that the true nonperturbative answer to the doubly-scaled unitary matrix model is given by the {\it unique}, real solution 
to Painlev\'e II with asymptotic behaviors (\ref{plusas}) and (\ref{minusas}). This is the Hastings--McLeod solution, which we will 
denote by $u_{\rm HM}$. This solution can be found by numerical integration, and we display it in \figref{HMfig}. 
On the other hand, we also know that 
any real solution to Painlev\'e II with the asympotics (\ref{plusas}) is of the form (\ref{realsol}). Therefore, there must be a real value of $C$, 
$C_{\rm HM}$, for which 
\be
u^{\IR}(\kappa;C_{\rm HM})=u_{\rm HM}(\kappa).
\ee
It turns out that this value is just $C_{\rm HM}=0$. 
One way to see this is to notice that the one-parameter family $u^{\IR}(\kappa;C)$ is the family of 
{\it tronqu\'ee} solutions studied by Boutroux \cite{painlevet}. In this 
family there is one single member with the asymptotics (\ref{minusas}), which is the Hastings--McLeod solution. The value 
$C_{\rm HM}=0$ can then be obtained by comparing the structure of $u^{\IR}(\kappa;C)$ with the results of \cite{painlevet} for the Stokes parameter of the 
{\it tronqu\'ee} solutions.

\FIGURE[!ht]{
\leavevmode
\centering
\epsfysize=4cm
 \leavevmode
    \hspace{3cm}\epsfbox{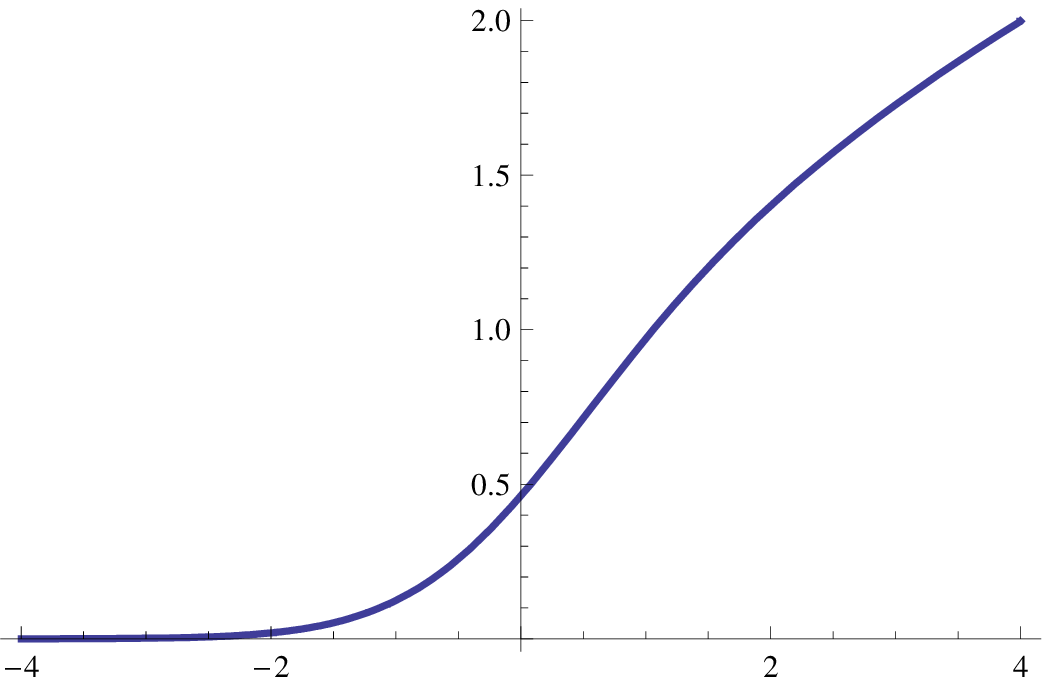}\hspace{3cm}
\caption{The Hastings--McLeod solution to Painlev\'e II. As $\kappa \rightarrow \infty$ it asymptotes $\sqrt{\kappa}$, while as $\kappa \rightarrow -\infty$ it 
decays exponentially like $\exp(-2 {\sqrt{2}} (-\kappa)^{3/2}/3)$.}
\label{HMfig}
}

We conclude that the Hastings--McLeod solution, which is the nonpertubative 
solution to the doubly-scaled unitary matrix model (and to many other models, see for example \cite{tw,pskpz}) has the trans-series expansion   
\be
\label{HMinst}
u_{\rm HM}(\kappa)=u^{\IR}(\kappa;0)={1\over 2} (u^{(0)}_+ +u^{(0)}_-) - {1\over 8} S^2 (u^{(2)}_+ +u^{(2)}_-) 
 +\cdots
 \ee
 for $\kappa$ sufficiently large. Notice that, according to this equation, the true nonperturbative solution of the problem has a 
 leading part, which is obtained by taking the real part of the Borel-resummed perturbative piece, and then it has instanton corrections 
 starting with {\it two} instantons. Recall that these instantons are obtained by taking the double-scaling limit of the unstable configurations 
 where eigenvalues of the unitary matrix 
 sit at the maximum of the potential $\theta=\pi$. Therefore, these configurations {\it do contribute} to the 
 physical answer for the partition function and have to be taken into account, as it was emphasized in \cite{mmss}.  

It is interesting to verify the relation (\ref{HMinst}) numerically. In this test we really need numerical precision, since the second term in (\ref{HMinst}) comes from a two-instanton correction 
and is very small. The most delicate part is to obtain accurate results for the numerical integration of the Painlev\'e II 
differential equation which gives $u_{\rm HM}(\kappa)$. 
Because of this we have relied on the results of \cite{pskpz}. In table \ref{table:solution} we compare the difference 
\be
 u_{\rm HM}(\kappa)-{\rm Re}\, u^{(0)}_+(\kappa)
 \ee
 to the two-instanton effect
 \be
 -{S^2 \over 4} \,{\rm Re} \, u_+^{(2)}(\kappa), 
 \ee
 and we evaluate it at different integer points $\bar \kappa=\kappa/2^{1\over 3}$. 
 The factor $2^{1/3}$ has been introduced in order to use the results of \cite{pskpz}, who obtain $u_{\rm HM}$ with a different normalization: 
 \be
u_{\rm HM}^{\rm ours}(\kappa) =-2^{1\over 3} u_{\rm HM}^{\rm theirs}(-2^{1\over 3} \kappa). 
\ee
 Again, the results we display in table \ref{table:solution} are exact at the precision we have used (the results of \cite{pskpz} have a precision of 16 digits, and this leads to the decrease in sensitivity in the table data as $\kappa$ grows). The agreement is excellent, and the differences between the two quantities 
 should be attributed to higher instanton corrections (in fact, as one can see from (\ref{manifreal}), these differences are four-instanton effects). 
 Our numerical results confirm indeed that $C_{\rm HM}=0$.

\TABLE[ht]{
\centering 
\hspace{3cm}\begin{tabular}{|c || c | c|} 
\hline 
$\bar \kappa$ & $ u_{\rm HM}(\kappa)-{\rm Re}\, u^{(0)}_+(\kappa)$ & $-S^2  \,{\rm Re} \, u_+^{(2)}(\kappa)/4 $  \\ [0.5ex] 
\hline 
$2$ & $0.000043768$ & $0.000043765$ \\ 
$3 $&$3.822644 \cdot 10^{-7}$ &$ 3.822642 \cdot 10^{-7}$ \\
$4$& $1.542393 \cdot 10^{-9} $ & $ 1.542393 \cdot 10^{-9}$ \\
$5$ & $ 3.176\cdot 10^{-12} $ & $3.176\cdot 10^{-12} $ \\
\hline 
\end{tabular}\hspace{3cm}
\caption{The left column of this table shows the difference between the Hastings--McLeod solution $u_{\rm HM}$ to Painlev\'e II, 
and the real part of the Borel-resummed perturbative series 
${\rm Re}\, u^{(0)}_+(\kappa)$, for various values of $\bar \kappa=\kappa/2^{1\over 3}$. This should be equal, at leading order, to the two-instanton effect $-S^2  \,{\rm Re} \, u_+^{(2)}(\kappa)/4 $, which 
we show in the right column. The agreement is excellent, and the difference for small $\bar \kappa$ is due to higher instanton corrections.} 
\label{table:solution} 
}

To summarize, we have shown that the formal trans-series solution to Painlev\'e II can be appropriately resummed to 
obtain a real, one-parameter family of true solutions with exponentially suppressed corrections due to multi-instantons and  
where nonperturbative ambiguities cancel. The parameter can be fixed by using further information from the nonperturbative result, and in particular we 
can reconstruct the Hastings--McLeod solution at least when $\kappa$ is sufficiently large. This gives a semi-classical expansion 
of this solution which includes instanton corrections.   

An interesting spinoff of our discussion is that the real part of the Borel-resummed perturbative series does {\it not} give the 
exact nonperturbative result, since there are higher instanton corrections in (\ref{HMinst}) starting at the two-instanton level. 
This has been also pointed out for the case of 
the double-well potential in quantum mechanics, where the multi-instanton corrections play a crucial role in reconstructing the exact 
answer for the energies \cite{zj, zjj}. In our case, this is intimately related to the nonlinearity of the 
differential equation encoding the exact answer: although the Borel-resummed series $u^{(0)}_{\pm}$ solve Painlev\'e II, their sum does not. 
 However, in situations where the physical answer to a resummation problem is known to be real, 
it is often assumed that one should simply take the real part of the Borel-resummed perturbative series 
(see \cite{john,raczka}). Our example, as well as the example of the double-well, 
show that this is not necessarily the right answer, and that further nonperturbative corrections are needed. 
A case where the real part of the Borel-resummed series is known to be the exact result appears in \cite{fateev}, but the relevant quantity 
studied in that paper satisfies a {\it linear} differential equation, therefore the sum of $u^{(0)}_{+}$ and $u^{(0)}_{-}$ 
is still a solution and higher corrections are absent. 

\subsection{Instantons and nonperturbative definition in the unitary matrix model}

The analysis of the previous section has produced a trans-series, formal solution for the free energy of the GWW model in the 
weakly coupled phase, of the form (\ref{fullf}). 
\be
\CF(t,g_s) =g_s^2 F(t, g_s) = \sum_{\ell=0}^{\infty}C^{\ell}  \CF^{(\ell)}(t,g_s), 
\ee
where we write
\be
\ba
\CF^{(0)}(t,g_s) &=\sum_{g=0}^{\infty} F_g(t) g_s^{2g}, \\
\CF^{(\ell)}(t,g_s) &=g_s^{2} \re^{-{\ell  A(t)\over g_s}}F_{\ell,1}(t) \varphi^{(\ell)} (t, g_s), \quad \ell \ge 1.
\ea
\ee
and 
\be
\varphi^{(\ell)}(t, g_s) =\sum_{n=0}^{\infty} F_{\ell,n+1}(t) g_s^n. 
\ee
The $g_s$ expansions involved here are factorially divergent series, therefore it is natural to consider as well the lateral Borel resummations
\be
\CF_{\pm} (t,g_s;C_{\pm}) =  \sum_{\ell=0}^{\infty}C_{\pm}^{\ell}  \CF^{(\ell)}_{\pm} (t,g_s), 
\ee
where
\be
\CF^{(\ell)}_{\pm} (t,g_s) =g_s^{2} \re^{-\ell  A(t)/g_s} F_{\ell,1}(t) \varphi^{(\ell)}_{\pm} (t, g_s), \quad \ell \ge 1.
\ee
Although our framework is slightly different from the one considered in the literature, it is known that some difference equations 
satisfy the same resurgence properties as differential equations (see for example \cite{braaksma}). It is then natural to conjecture that the resurgence relation (\ref{resurgence}) 
becomes, in the setting of the full matrix model, 
\be
\label{fullresurgence}
\CF_{+} (t,g_s;C)=\CF_{-} (t,g_s;C-S^{\rm w}), 
\ee
where $S^{\rm w}$ is the Stokes multiplier (\ref{weaks}). This leads to a cancellation of nonperturbative ambiguities similar to the one considered 
before. At the one instanton level, we have
\be
\label{fullresdet}
\CF_+^{(0)} -\CF^{(0)}_-={S^{\rm w}\over 2}(\CF^{(1)}_++\CF^{(1)}_-) + \cdots.
\ee
We have tested this relation numerically by using Pad\'e--Borel resummation of the sequences $F_g(t)$ (up to $g=25$) and 
$F_{1,n}$ (up to $n=15$). We found very good agreement. We show some results for $t=1/2$ and various values of $N$ in table \ref{table:ucan}. 
Notice that in these calculations we have less precision since we computed fewer terms in the series, and we are not able to resolve higher instanton 
corrections. 

\TABLE[ht]{
\centering 
\hspace{3cm} \begin{tabular}{|c || c | c|} 
\hline 
$N$ & $ {\rm Im} \, \CF_+^{(0)}(t,g_s)$ & $-\ri S  \,{\rm Re} \, \CF_+^{(1)}(t,g_s)/2 $  \\ [0.5ex] 
\hline 
$2$ & $0.000303$ & $0.000303$ \\ 
$3 $&$ 0.0000491$ &$ 0.0000491$ \\
$4 $& $7.440 \cdot 10^{-6}$ & $7.440 \cdot 10^{-6}$ \\
$5$ & $1.5219 \cdot 10^{-6} $ & $1.5219 \cdot 10^{-6} $ \\
$10$ & $1.4578  \cdot 10^{-9}$ &  $1.4578 \cdot 10^{-9}$ \\ [1ex] 
\hline 
\end{tabular}\hspace{3cm}
\caption{Cancellation of nonperturbative ambiguities in the GWW model for $t=1/2$ and various values of $N$.} 
\label{table:ucan} 
}

Assuming the resurgence relation (\ref{fullresurgence}) to hold, we can produce a one-parameter family 
of real solutions to the 
difference equations by taking
\be
\label{realf}
\CF^{\IR}(t,g_s;C)=\CF_{+} (t,g_s;C+S^{\rm w}/2)=\CF_{-} (t,g_s;C-S^{\rm w}/2), \qquad C\in \IR.
\ee

We now compare the resummation of the $g_s$ expansion and its 
multi-instanton corrections to the exact nonperturbative answer. Based on the analysis of the 
double-scaling limit, we expect that the resummed solution to the difference equation (\ref{realf}) with $C=0$
\be
\label{exactas}
\CF^{\IR}(t,g_s;0)={1\over 2} (\CF_+^{(0)} +\CF^{(0)}_-)-{(S^{\rm w})^2\over 8}(\CF^{(2)}_++\CF^{(2)}_-) + \cdots
\ee
is the exact, semiclassical expansion of the full nonperturbative free energy (as long as $A(z)/g_s$ is big enough). 
In order to test this relation, we have to determine the nonperturbative free energy of the GWW model. Since in the computation of the 
perturbative part we have subtracted the Gaussian free energy, we have
\be
\CF^{\rm np} (N, g_s)=  g_s^2 \log \prod_{i=0}^{N-1} h_i  -g_s^2 F_G, 
\ee
where the Gaussian free energy can be computed exactly in terms of the Barnes function, 
\be
 F_G={N^2 \over 2} \log g_s+ \log \biggl[ { G_2(N+1) \over (2\pi)^{N\over 2} }\biggr]. 
 \ee
The product of the $h_i$ can be computed as a Toeplitz determinant \cite{wadia,bg}
 \be
\log \prod_{i=0}^{N-1} h_i={\rm det} \Bigl[ I_{k-l}(1/g_s) \Bigr]_{k,l=1, \cdots, N}
\ee
where
\be
I_n(z) =I_{-n}(z) =\int_0^{2\pi} {\rd \theta \over 2 \pi} \re^{\ri n \theta + z \cos \theta}, 
\ee
are modified Bessel functions. We then find that the full nonperturbative answer for the free energy is 
\be
\label{exactgww}
\CF^{\rm np}(N, g_s)=g_s^2\Biggl\{ \log {\rm det} \Bigl[ I_{k-l}(1/g_s) \Bigr]_{k,l=1, \cdots, N} -\log \biggl[ { G_2(N+1) \over (2\pi)^{N\over 2} }\biggr]\Biggr\} -{g_s^2 N^2 \over 2}
\log g_s, 
\ee
which can be calculated exactly for any $N, g_s$. According to our results, this exact function of $N, g_s$ has an asymptotic expansion 
of the form (\ref{exactas}). 

As a partial verification of 
this statement, we have evaluated the first term in (\ref{exactas}), i.e. the real part of the Borel resummation of the $1/N$ expansion, and compared it 
to the exact expression (\ref{exactgww}) for various values of $N$, $t$. As in the test of (\ref{fullresdet}), we truncated the series at $g=25$. 
The agreement is excellent and the difference between both values 
should come from instanton corrections starting at two-instantons, as we checked in detail in the double-scaled model. In \figref{exgww} the continuous line represents ${\rm Re}\, \CF^{(0)}_+(t,g_s)$ for $t=1/2$ as a function of $1\le N\le 10$ (notice that, in the $1/N$ expansion, we can take $N$ to be a continuous variable, equal to $t/g_s$). The dots represent the exact result (\ref{exactgww}) for the integers $N=1, \cdots, 10$. As $N\rightarrow \infty$, both 
quantities asymptote the planar limit $F_0(t)=t=1/2$. Notice that this test does not verify that the right value of $C$ in (\ref{realf}) is indeed $C=0$, but this value is required in order to match 
the results in the double-scaling limit. 

\FIGURE[ht]{
    \centering
     \epsfxsize=.5 \textwidth
    \leavevmode
    \hspace{3cm}\epsfbox{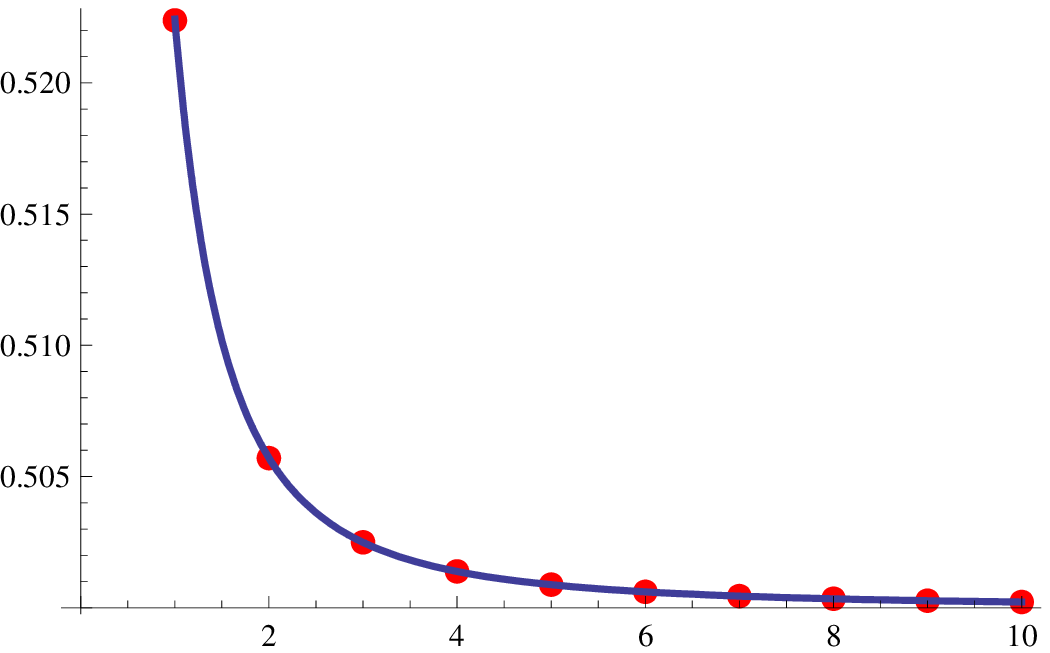}\hspace{3cm}
\caption{Comparison of the exact nonperturbative result $\CF^{\rm np}(N, g_s)$ (dots) 
with the first term in the Borel resummed trans-series expansion 
${\rm Re}\, \CF^{(0)}_+(t,g_s)$ (continuous line). The horizontal axis represents $N$, 
the rank of the unitary matrix. The 't Hooft parameter is fixed to $t=1/2$. 
Notice that both results are asymptotic, as $N \rightarrow \infty$, to the planar limit $F_0(t)=t=1/2$.}
\label{exgww}
}

The question of how accurate is the $1/N$ expansion in order to reproduce the full nonperturbative answer has been an important one since 
this expansion was first formulated. In the particular case of the GWW model, a preliminary investigation of this issue was already performed 
in the seminal paper by Wadia \cite{wadia}. We can now summarize our results, which give an extremely detailed answer to this question in the 
weakly coupled phase:

\begin{itemize}

\item The $1/N$ expansion of the unitary matrix model, in the weak coupling phase, is not Borel summable, and diverges factorially like 
$(2g)!$. Its large order behavior is governed by a one-instanton amplitude. 

\item This $1/N$ expansion can however be resummed by using Borel transforms and lateral resummations. 
Ambiguities coming from different contour prescriptions cancel 
against the ambiguity in the one-instanton contribution. 

\item The exact nonperturbative answer for the model is given by the real part of the Borel-resummed $1/N$ expansion, plus 
an infinite series of exponentially suppresed corrections. These corrections can be computed as Borel-resummed instanton expansions, 
and they start at {\it two} instantons. 

\end{itemize}

The importance of instantons in the weakly coupled phase of the unitary GWW model was pointed out by Neuberger 
in \cite{neuberger}, although no computational scheme was provided there to derive them. Their importance 
is closely related to the existence of 
a phase transition, since instanton contributions will become more and more 
relevant as we approach the transition point, where their action $A(t)$ vanishes. In particular, the trans-series expansion (\ref{exactas}) will break down 
at $t=1$, and in this sense we can say that the phase transition in the free energy is triggered by instantons which are no longer 
suppressed exponentially \cite{neuberger,gma}. 

Although we have focused on the weakly coupled phase, it is in principle possible to obtain a similar convergent instanton expansion in the strongly coupled phase.

\subsection{The Hermitian case}

We have seen that, starting from the formal multi-instanton series, we can form a one-parameter family of solutions 
to the original differential or difference equations by considering the Borel resummations. The remaining parameter is fixed by 
nonperturbative input, and we did that in the case of Painlev\'e II/unitary matrix model.

\FIGURE[!ht]{\label{cubicplane}
\centering \leavevmode
\epsfysize=5cm
\epsfbox{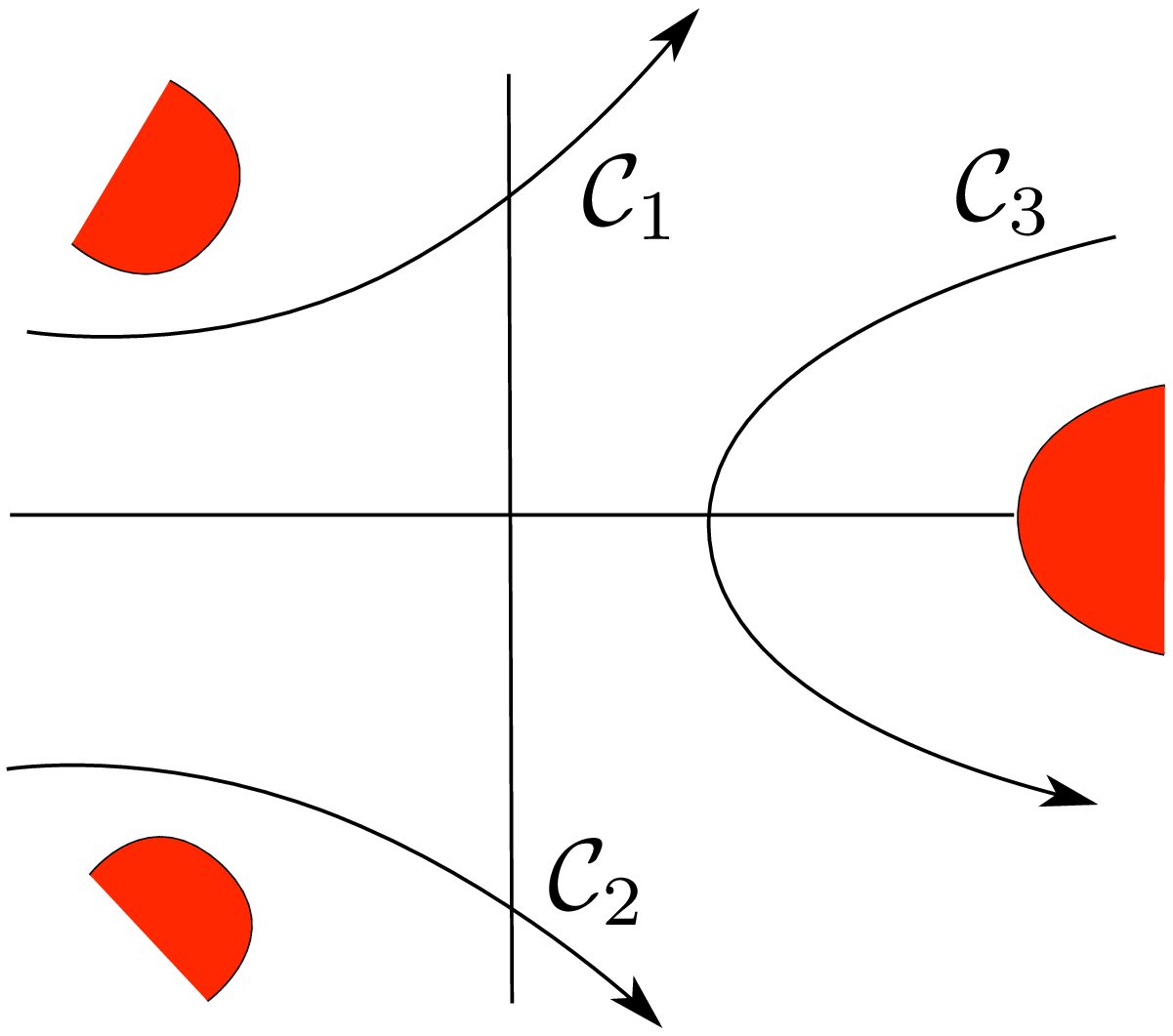}
\caption{Three integration contours where the cubic matrix integral defined by (\ref{cubicd}) is convergent.}}

In the case of Hermitian matrix models, and as we remarked in section 3, we can obtain convergent matrix integrals by suitably choosing the integration 
contours for the eigenvalues \cite{david,davidvacua}. Using the methods of section 3 we can obtain a formal trans-series expression for this matrix integral, 
and by using Borel resummation we can recover (at least for a one-dimensional 
submanifold of the parameter space) the original, convergent matrix integral. Let us sketch how this would work in 
the case of a cubic matrix model with potential
\be
\label{cubicd}
V(z)= z -{1\over 3} z^3. 
\ee
We can define a convergent matrix integral as in \cite{davidvacua} by choosing integration contours in the complex plane where the potential 
decreases at infinity. Three such contours are shown in \figref{cubicplane}. Of course, these 
contours are linearly dependent, since
\be
\CC_2=\CC_1 + \CC_3. 
\ee
The most general, convergent matrix integral obtained in this way is of the form (\ref{sumz})
\be
\label{zgam}
Z= \sum_{N_1+N_2=N} \zeta_1^{N_1} \zeta_2^{N_2} Z(N_1,N_2).
\ee
where $N_1$ eigenvalues are integrated along the contour 
$\CC_1$, and $N_2$ eigenvalues are integrated along the contour $\CC_2$. As noticed in \cite{davidvacua}, if we choose 
\be
\label{realzeta}
\zeta_{1,2}={1\over 2} \pm \ri \theta
\ee
the resulting integral is real, since the contours $\CC_{1,2}$ are complex conjugate. 

On the other hand, with the procedure explained in section 3, we can obtain for the free energy of the cubic matrix model a trans-series expansion
\be
F(t,g_s)=\sum_{\ell=0}^{\infty} C^{\ell} F^{(\ell)}(t, g_s),
\ee
where $F^{(\ell)}(t,g_s)$ is the $\ell$-instanton solution. From here, through lateral Borel resummations, we can construct a true, real, one parameter solution to the relevant difference equation
\be
\label{realcubic}
F^{\IR}(t,g_s;C)=\sum_{\ell=0}^{\infty} (C-S/2)^{\ell} F^{(\ell)}_+(t, g_s)
\ee
in analogy with (\ref{realsol}). We have to match now the family (\ref{realcubic}) to the family of partition functions (\ref{zgam}) with coefficients 
(\ref{realzeta}). This can be done by requiring that the one-instanton term in 
(\ref{realcubic}) matches the one-instanton term for (\ref{zgam}) as computed 
in \cite{davidvacua}. This is straightforward and one obtains in this way
\be
C =\ri \theta S,
\ee
which is real since $S$ is pure imaginary. We conclude that, for this choice of $C$, (\ref{realcubic}) gives a convergent series expansion for the logarithm of the 
nonperturbative 
answer (\ref{zgam}) where the $\zeta_{1,2}$ are given by (\ref{realzeta}). As a further check of this relation, 
notice that, in the double scaling limit, (\ref{realcubic}) becomes the solution to 
Painlev\'e I given by 
\be
u_+\biggl(\kappa; \Bigl( -{1\over 2} + \ri \theta\Bigr) S\biggr). 
\ee
If $\theta$ is not real, these are complex solutions, and in particular if $\theta =\mp \ri /2$ we should obtain the 
so-called triply truncated solutions of Painlev\'e I \cite{davidvacua}. Using (\ref{resurgence}) we see that these triply truncated solutions correspond, respectively, to 
\be
u_+(\kappa;0), \qquad u_-(\kappa;0), 
\ee
i.e. the lateral Borel resummations above and below the real axis,  which give indeed representations of the triply truncated solutions (see for example section 
5.6 of \cite{davidreview}).

\sectiono{Nonperturbative effects in topological string theory}

\subsection{General picture}

The free energy of topological string theory on a Calabi--Yau threefold $X$ 
can be regarded as a quantum mechanical system with two different Planck constants. 
The {\it worldsheet} Planck constant is given by the square of the string length,
\be
\hbar_{\rm ws} =l_s^2. 
\ee
For a fixed genus, the free energy $F_g(t)$ near the large radius limit has an expansion of the form 
\be
F_g(t)=\sum_{n_i \ge 0}N_{g,n}\, \re^{-n \cdot t /\hbar_{\rm ws}}.
\ee
Here, the sum over $n_i$, $i=1, \cdots, b_2(X)$, 
is a sum over topological sectors, or equivalently, over worldsheet instanton numbers, and $n \cdot t =\sum_{i=1}^{b_2(X)} n_i t_i$ 
can be interpreted in the A model as 
the action of a worldsheet instanton with instanton numbers $n_i$, and it depends on the K\"ahler parameters $t_i$ of $X$. 
In principle, we should expect a perturbative expansion in $\hbar_{\rm ws}$ around the instanton, but 
the presence of worldsheet $\CN=(2,2)$ supersymmetry implies that the only nonvanishing term in this series occurs at one-loop, and gives 
the Gromov--Witten invariant $N_{g,n}$. 

There is however a second, {\it spacetime} Planck constant which is the string coupling constant
\be
\hbar_{\rm st}=g_s. 
\ee
Indeed, the (perturbative) free energy is given by a series of the form
\be
F^{(0)}(t,g_s) =\sum_{g=0}^{\infty} F_g(t) g_s^{2g-2}.
\ee
It is then natural to conjecture that the {\it full} free energy of topological string theory should be in general 
a trans-series expansion depending on two small parameters, namely 
\be
\label{twop}
g_s, \qquad \re^{-A(t)/g_s}
\ee
where $A(t)$ is the action of a {\it spacetime} instanton, i.e.
\be
\label{transts}
F(t,g_s) =\sum_{\ell=0}^{\infty} F^{(\ell)}(t,g_s), \qquad F^{(\ell)}(t,g_s)\sim \re^{-\ell A(t)/g_s}, \quad \ell\ge 1.
\ee
Here we have assumed for simplicity that all spacetime instantons are classified by a single instanton number $\ell$, but of 
course there can be more general situations. 

This conjecture was put forward in \cite{msw} by using the connection between topological strings and matrix models, and some simple 
examples of topological string theory were studied there from this point of view. These examples are topological string theory on the toric Calabi--Yau 
\be
\label{xp}
X_p=\CO(-p) \oplus \CO(p-2) \rightarrow \IP^1
\ee
and its $p\rightarrow \infty$ limit, which can be interpreted as a theory of simple Hurwitz numbers. It was proposed in \cite{mm} 
that these topological string models models can be described in terms of spectral curves akin to those appearing 
in matrix models, and this was shown to be the case in \cite{eynardproof}. 
This makes possible to calculate explicitly 
the spacetime instanton action $A(t)$, as well as the first few terms of the one-instanton contribution 
to the free energy $F^{(1)}(t,g_s)$, by using saddle-point, matrix model techniques. 
We will now provide further evidence for this general conjecture by analyzing simple topological string models with the techniques and ideas developed above. 

\subsection{A toy model}
In the previous sections we have shown that multi-instanton series can be obtained, in the case of matrix models, by finding trans-series solutions to the difference 
equations that describe the model. In general, it is not known if the free energies of topological string models are described by differential or difference equations. Here we point out that the Hurwitz model studied in \cite{msw} can be described by a difference equation \cite{p} 
which admits a trans-series solution. Therefore, the total free energy of the Hurwitz model is indeed of the form (\ref{transts}). We will also  
verify that this solution reproduces the one-instanton effects computed in \cite{msw}. 

The Hurwitz model is defined, in down-to-earth terms, by a partition function of the form 
\be
Z(t_H,g_H)=\sum_{g\ge 0} g_H^{2g-2} \sum_{d\ge 0} {H_{g,d}^{\IP^1} (1^d)  \over (2g-2+2d)!}Q^d,
\ee
where $Q=\re^{-t_H}$ and $g_H$ can be regarded as formal parameters keeping track of the degree and the genus, respectively, and 
$H_{g,d}^{\IP^1} (1^d)$ is a simple Hurwitz number counting degree $d$ covering maps of $\IP^1$, with simple branch points only , and by Riemann surfaces 
of genus $g$ (see \cite{msw} for explicit expressions). The free energy $\log\, Z$ describes connected, simple 
Hurwitz numbers $H_{g,d}^{\IP^1} (1^d)^{\bullet}$,
\be\label{freehurwitz}
F = \log Z = \sum_{g\ge 0} g_H^{2g-2} \sum_{d\ge 0}  {H_{g,d}^{\IP^1} (1^d)^{\bullet} \over (2g-2+2d)!}Q^d,
\ee
\noindent
and it has the genus expansion
\be\label{genushurwitz}
F(g_H,t_H) = \sum_{g=0}^{\infty} g_H^{2g-2} F_g(Q).
\ee

This theory is in fact a topological string theory in disguise. It can be realized as a special limit of topological string theory on certain toric Calabi--Yau manifolds, 
see for example \cite{bp,italy} for detailed derivations. It was conjectured in \cite{mm} and proved in \cite{eynardproof} that Hurwitz theory can be described in terms of 
matrix integrals, and this in turn was used in \cite{msw} to compute the one-instanton contribution to the perturbative free energy (\ref{genushurwitz}). In particular, it was found 
in \cite{msw} that the instanton action is given by
\be
\label{mmin}
A(t_H)=2w \Bigl( \chi + \cosh(w) \chi^{1\over 2} -2\Bigr),
\ee
In this equation, the dependence on $t_H$ occurs through the variable $\chi$ defined by
\be
\label{chiQ}
\chi\re^{-\chi} =\re^{-t_H},
\ee
and $w$ is defined by the implicit equation
\be
{w\over \sinh(w)} =\chi^{1\over 2}.
\ee

As shown in \cite{p}, the free energy of Hurwitz theory satisfies a difference equation of the Toda type, 
\be
\label{toda}
\exp \Bigl( F(t_H+g_H) +F(t_H-g_H)-2 F(t_H)\Bigr)= g_H^2 \re^{t_H} \partial_{t_H}^2 F(t_H,g_H). 
\ee
As we did in section 3, we can try to solve this equation with a trans-series ansatz of the form (\ref{fullf}). Doing this one immediately obtains 
the following equation for the one-instanton amplitude, 
\be
\label{hurwitzone}
\exp \Bigl( \Delta_{g_H} F^{(0)}(t_H)\Bigr) \Delta_{g_H} F^{(1)}(t_H) = g_H^2 \re^{t_H} \partial_{t_H}^2 F^{(1)},
\ee
where we have written
\be
\Delta_{h} f(t)=
f(t+h) +f (t-h)-2 f(t)
\ee
to denote the discrete Laplace operator with step $h$. The first term in the expansion of (\ref{hurwitzone}) in 
powers of $g_H$ gives an equation for $A'(t_H)$, 
\be
\label{hurins}
2\Bigl[ \cosh (A'(t_H)) -1\Bigr]=\re^{t_H-\chi} (A'(t_H))^2, 
\ee
where we used that
\be
\partial_{t_H}^2 F^{(0)}_0(t_H)= \chi.
\ee
One can check that the function $A(t_H)$ defined implicitly by (\ref{hurins}) coincides with 
the instanton action (\ref{mmin}) computed by the matrix model (we tested this by 
expanding both quantities around the critical point of the model at $\chi=1$). Of course, it is straightforward to use (\ref{toda}) 
to derive a full trans-series solution for the free energy. By the arguments already explained above, one can 
perform if needed Borel resummations to obtain a one-parameter family of true solutions. We expect from the general arguments 
explained in section 4 that all solutions to the Toda-like equation (\ref{toda}) with the asymptotics fixed by the perturbative 
expansion are described by this one-parameter family. Therefore, a nonperturbative completion of the 
theory should be equivalent to fixing a value for this parameter.

\subsection{Holographic description and nonperturbative effects}

In the example considered above, as well as in the more general example of topological string theory on local curves studied in \cite{msw}, 
one can compute nonperturbative effects by using a matrix model dual description. On the other hand, there is strong evidence \cite{mm,bkmp}
that the closed and open amplitudes of topological string theory on a toric Calabi--Yau threefold can be described in terms of recursion relations on a 
spectral curve typical from matrix models \cite{eo}. It is then natural to expect that, in the same way that the full matrix model partition function involves a trans-series 
expansion obtained by summing over all instanton sectors (i.e. over all filling fractions), as in (\ref{sumz}), the full partition function for topological string 
theory on a toric Calabi--Yau threefold will involve such a sum over spacetime instanton sectors, as we have conjectured above. 

A partial verification of this expectation, beyond the simple toy model considered before, comes from looking at topological strings with large $N$ Chern--Simons duals. 
In particular, topological string theory on $A_{p-1}$ fibrations over $\IP^1$ is conjectured to be dual to 
Chern--Simons theory on the lens space $\IS^3/\IZ_p$ \cite{akmv}, and 
some detailed evidence for this was obtained in \cite{akmv,hy,hoy}. 
In this case, near the orbifold point in moduli space which is dual to the Gaussian point of Chern--Simons theory, the matrix model realization can be made explicit by using the matrix integral representation of Chern--Simons theory found in \cite{mmint}. One finds 
that the {\it perturbative} topological string partition function, near the orbifold point, is given by a matrix integral similar to (\ref{genz})
\be
\label{csmm}
\ba
&Z(t_1, \cdots, t_p)=Z(N_1, \cdots, N_p) \\
&= {1\over N_1! \cdots N_p!} \int_{\lambda^{(1)}_{i_1} \in \CC_1} \cdots \int_{\lambda^{(p)}_{i_p} \in \CC_p}  
\prod_{i=1}^N {\rd\lambda_i \over 2 \pi}\, \prod_{i<j}\biggl( 2 \sinh {\lambda_i -\lambda_j \over 2} \biggr)^2 \re^{-{1\over 2 g_s} \sum_{i=1}^N 
(\lambda_i -\lambda_i^*)^2}.
\ea
\ee
In this equation, 
\be
\lambda_i^*={2\pi \ri \over p} \Bigl( \overbrace{0, \cdots, 0}^{N_1}, \overbrace{ 1, \cdots, 1}^{N_2}, \cdots, \overbrace{ 
p-1 , \cdots, p-1}^{N_p}\Bigr),
\ee
the contour $\CC_k$ passes through the point $2\pi \ri (k-1)/p$, and the K\"ahler parameters of the Calabi--Yau 
$t_i$ are identified with the 't Hooft parameters 
$g_s N_i$. The {\it nonperturbative} answer, which is the Chern--Simons partition function on $\IS^3/\IZ_p$, is 
given by \cite{mmint,akmv} 
\be
\label{csmatrix}
Z_{\rm CS}(N, g_s) =\sum_{N_1+\cdots+N_p=N} \zeta_1^{N_1} \cdots \zeta_p^{N_p} Z(N_1, \cdots, N_p), 
\ee
where 
\be
\zeta_j=\exp\Bigl( {\pi \ri \hat k \over p} (j-1)^2 \Bigr), \quad j=1, \cdots, p
\ee
and $\hat k=k+N$ is the shifted coupling constant of Chern--Simons theory, which is related to the string coupling constant by 
\be
g_s= {2\pi \ri \over p \hat k}. 
\ee
Notice that, in Chern--Simons theory, $\hat k$ is an integer. 

The expression (\ref{csmatrix}) is precisely of the form (\ref{sumz}). Therefore, in topological string theories on toric Calabi--Yau threefolds with large $N$ duals, 
the nonperturbative definition of the topological string partition function, which can be read from the Chern--Simons 
gauge theory dual, involves a sum over all the instanton sectors of the matrix model (\ref{csmm}). 
This provides a further confirmation of the conjectural structure of the 
full topological string partition function as involving a sum over spacetime instanton sectors. 

In \cite{eynard}, Eynard has pointed out that the sum over multi-instantons in a matrix model is independent on the choice of filling fractions, 
and therefore should be background independent. Indeed, since in an expression like (\ref{sumz}) and (\ref{csmatrix}) we sum over all 
possible backgrounds $t_i=g_s N_i$, the final result should not depend on any particular choice of background. This is in contrast to the 
perturbative topological string free energy, where one has chosen a fixed, arbitrary background given by $t_i=g_s N_i$. Notice that, albeit the $t_i$ transform in a nontrivial way under the symplectic group acting on the periods of the Calabi--Yau, the total 't Hooft coupling $t=t_1 +\cdots +t_p$, which is the variable appearing in the l.h.s. of 
(\ref{csmatrix}), should be modular invariant for this picture to be consistent. For $p=2$ this can be checked by using the results of \cite{akmv}. As we have just seen, holographic duals 
force us to consider precisely the sum over instanton sectors as a natural nonperturbative definition of the full topological string partition function. This suggests 
that (\ref{csmatrix}) is a natural starting point to construct background independent topological string models.

\sectiono{Conclusions and open problems}

In this paper we have studied various aspects of nonperturbative effects in matrix models. First of all, we have developed techniques to compute 
multi-instanton amplitudes by finding trans-series solutions to the relevant difference equations, and studied in some detail both the Hermitian, 
quartic matrix model, and the unitary GWW model. These techniques give formal, asymptotic series, and by using results from the theory of 
resurgent functions and of exponential asymptotics, we spelled out in detail how to obtain convergent series which can then be used to find 
multi-instanton expansions of the true, nonperturbative matrix integrals. We illustrated this in the case of the unitary matrix model and the GWW model, and we clarified in this 
way some subtle aspects of the $1/N$ expansion which might be relevant in more complicated situations. 
Finally, following \cite{msw}, we argued that these trans-series instanton expansions should be also relevant in 
topological string theory, and we gave some 
pieces of evidence for this. In particular, we showed that in topological strings with both a matrix description and a 
holographic Chern--Simons dual, the nonperturbative definition in terms of the 
gauge theory partition function indeed forces us to consider all the instanton sectors of the matrix model. 

There are many aspects of the paper that should be further clarified and extended. We end with a list of open problems which we find interesting. 

\begin{itemize} 

\item The strategy followed here to study Painlev\'e II can be also used to study nonperturbative effects in the $(p,q)$ minimal string. In particular, the non-unitary models 
(like the $(2,5)$ model that describes the Yang--Lee singularity) are well-defined nonperturbatively \cite{bmp} and one could ``unfold" the semi-classical content 
of the exact answer by using the approach based on trans-series solutions. Another closely related example is the weak coupling phase of the minimal superstring with flux, which 
is described by a close cousin of Painlev\'e II \cite{kms,ss}. The only nontrivial information, namely the value of the Stokes parameter as a function of the flux, 
can be inferred from the results in \cite{painlevet}. 

\item The instanton effects computed for the weakly coupled phase of the GWW model should be inherited in the 
multi-trace unitary models used to describe (super) Yang--Mills theory on $\IS^3 \times \IS^1$. These effects are of order $\CO(\re^{-N})$ 
and they might provide tractable gauge theory duals to 
D-brane effects on the string side. 

\item Another model that one could study with these techniques is the nonperturbative completion of 2d gravity proposed in 
\cite{johnson}, which is described by a different string equation yet has the same asymptotics than Painlev\'e I. Since the perturbative 
asymptotics determines the Stokes parameter and the one-instanton amplitude, the formal trans-series extension of the solution proposed in \cite{johnson} 
must share many properties with the trans-series solution to Painlev\'e I, and it would be interesting to understand their relation in more detail. 

\item Although the results we use for Painlev\'e II are corollaries of more general results in the theory or resurgent functions and of exponential asymptotics, 
the results we have presented 
for the full matrix model have not been established rigorously. It would be interesting to show, by extending known results and techniques, 
that the properties we have assumed and tested numerically indeed hold for the difference equations characterizing matrix models. This is potentially a very 
rich arena, since all the relevant quantities in the trans-series asymptotics, 
like the instanton action, depend now on parameters (the 't Hooft coupling and the coupling constants of the model), and we will have a very rich situation in which the analyticity structure (for example, 
the location of the poles of the Borel transform) changes as we move in parameter space.

\item It seems very likely that the topological string theory on the Calabi--Yau threefold $X_p$ defined in (\ref{xp}) is also described by a difference equation which generalizes 
(\ref{toda}). It would be interesting to find such an equation and use it to obtain trans-series solutions. Of course, it would be even more interesting to 
find explicit difference equations for the topological string partition function on other Calabi--Yau targets, or to translate the matrix model results of \cite{mm,bkmp} in such a 
framework. 

\item As we mentioned in section 3, general instanton amplitudes are closely related to multicut amplitudes, and further clarification of this relationship should be 
beneficial in the study of nonperturbative effects in matrix models. Results in this direction will appear in \cite{mswta}. 

\item As we explained in the last section, for 
topological strings with matrix model as well as holographic duals, the idea of completing the perturbative topological string partition function by 
adding instanton sectors of the matrix model is not only reasonable; it is in fact {\it imposed} to us by the holographic dual. It would be very important to clarify these  nonperturbative effects and to understand their implications for background independence in string theory, as pointed out in \cite{eynard}, and more generally for large $N$ dualities as a whole.
 
\end{itemize}
  
\section*{Acknowledgments}
I would like to thank Fran\c cois David, Bertrand Eynard, 
Stavros Garoufalidis, Ulrich Jentschura, Sara Pasquetti, Marlene Weiss and Peter Wittwer for useful discussions and/or correspondence. I'm particularly indebted to 
Jean--Pierre Eckmann, Ricardo Schiappa and Spenta Wadia for their comments and for a detailed reading of the manuscript. 






\bibliographystyle{plain}

\begin{thebibliography}{10}

\bibitem{akmv}
  M.~Aganagic, A.~Klemm, M.~Mari\~no and C.~Vafa, ``Matrix model as a mirror of Chern-Simons theory,''
  JHEP {\bf 0402}, 010 (2004)
  [arXiv:hep-th/0211098].
  
\bibitem{akk}
S.~Y.~Alexandrov, V.~A.~Kazakov and D.~Kutasov, ``Non-perturbative effects in matrix models and D-branes,''
  JHEP {\bf 0309}, 057 (2003)
  [arXiv:hep-th/0306177].
  
 \bibitem{alvarez}
G. \'Alvarez, ``Coupling-constant behavior of the resonances of the cubic anharmonic oscillator," Phys. Rev. A {\bf 37},  4079 (1988). 
  
 \bibitem{abw} 
T.~Azuma, P.~Basu and S.~R.~Wadia, ``Monte Carlo Studies of the GWW Phase Transition in Large-N Gauge Theories,''
  Phys.\ Lett.\  B {\bf 659}, 676 (2008)
  [arXiv:0710.5873 [hep-th]].
 
\bibitem{bg}
I.~Bars and F.~Green, ``Complete Integration Of $U(N)$ Lattice Gauge Theory In A Large $N$ Limit,''
  Phys.\ Rev.\  D {\bf 20}, 3311 (1979).

\bibitem{bo}
C.M. Bender and S.A. Orszag, {\it Advanced mathematical methods for scientists and engineers}, Springer Verlag, 1999.

\bibitem{bessis}
D. Bessis, ``A New Method In The Combinatorics Of The Topological Expansion,''
Commun.\ Math.\ Phys.\  {\bf 69}, 147 (1979).

\bibitem{biz}
D.~Bessis, C.~Itzykson and J.~B.~Zuber,
``Quantum Field Theory Techniques In Graphical Enumeration,''
Adv.\ Appl.\ Math.\  {\bf 1}, 109 (1980).

\bibitem{bde}
G.~Bonnet, F.~David and B.~Eynard,
``Breakdown of universality in multi-cut matrix models,''
J.\ Phys.\ A {\bf 33}, 6739 (2000)
[arXiv:cond-mat/0003324].

\bibitem{bkmp}
 V.~Bouchard, A.~Klemm, M.~Mari\~no and S.~Pasquetti, ``Remodeling the B-model,''
  arXiv:0709.1453 [hep-th].
\bibitem{braaksma}
B. L. J. Braaksma, ``Transseries for a class of nonlinear difference equations," J. Differ. Equations Appl. {\bf 7} (2001) 717. 

\bibitem{bk}
E.~Br\'ezin and V.~A.~Kazakov,
``Exactly Solvable Field Theories Of Closed Strings,''
Phys.\ Lett.\ B {\bf 236}, 144 (1990) .

\bibitem{bmp}
E.~Br\'ezin, E.~Marinari and G.~Parisi, ``A Nonperturbative Ambiguity Free Solution Of A String Model,''
  Phys.\ Lett.\  B {\bf 242}, 35 (1990).
  
\bibitem{bp}
J.~Bryan and R.~Pandharipande, ``The local Gromov-Witten theory of
curves,'' [arXiv:math.AG/0411037.]

\bibitem{useful}
E.~Caliceti, M.~Meyer-Hermann, P.~Ribeca, A.~Surzhykov and U.~D.~Jentschura, ``From Useful Algorithms for Slowly Convergent Series to Physical Predictions Based on Divergent Perturbative Expansions,'' Phys. Rep. {\bf 446} (2007) 1 [arXiv:0707.1596 [physics.comp-ph]].

\bibitem{approche}
B. Candelpergher, J.C. Nosmas and F. Pham, {\it Approche de la r\'esurgence}, Hermann, Paris, 1993.

\bibitem{italy}
 N.~Caporaso, L.~Griguolo, M.~Mari\~no, S.~Pasquetti and D.~Seminara, ``Phase transitions, double-scaling limit, and topological strings,''
  Phys.\ Rev.\  D {\bf 75}, 046004 (2007)
  [arXiv:hep-th/0606120].
  
\bibitem{costin}
O. Costin, ``Exponential asymptotics, transseries, and generalized Borel summation for analytic, nonlinear, rank-one systems of ordinary differential equations," 
Internat. Math. Res. Notices {\bf 8} (1995) 377 [arXiv:math.CA/0608414]. 

\bibitem{costincostin}
O. Costin and R. Costin, ``On the formation of singularities of solutions of nonlinear differential systems in antistokes directions," 
Invent. Math. {\bf 145}, 425 (2001) [arXiv:math.CA/0202234].

\bibitem{cdm}
C.~Crnkovic, M.~R.~Douglas and G.~W.~Moore, ``Physical solutions for unitary matrix models,''
  Nucl.\ Phys.\  B {\bf 360}, 507 (1991).
 
 \bibitem{johnson}
  S.~Dalley, C.~V.~Johnson and T.~R.~Morris, ``Nonperturbative two-dimensional quantum gravity,''
  Nucl.\ Phys.\  B {\bf 368}, 655 (1992).
  
  \bibitem{davidren}
 F.~David, ``On The Ambiguity Of Composite Operators, IR Renormalons And The Status Of The Operator Product Expansion,''
  Nucl.\ Phys.\  B {\bf 234}, 237 (1984).
  
 \bibitem{david}
 F.~David,
  ``Phases Of The Large N Matrix Model And Nonperturbative Effects In 2-D Gravity,''
  Nucl.\ Phys.\ B {\bf 348}, 507 (1991).

\bibitem{davidvacua}
F.~David, ``Nonperturbative effects in matrix models and vacua of two-dimensional gravity,''
  Phys.\ Lett.\  B {\bf 302}, 403 (1993)
  [arXiv:hep-th/9212106].
  
  \bibitem{davidreview}
  F.~David, ``Simplicial quantum gravity and random lattices,''
  arXiv:hep-th/9303127.
 
  \bibitem{deift}
P. Deift, ``Universality for mathematical and physical systems," International Congress of Mathematicians. Vol. I, 125--152, Eur. Math. Soc., 
Z\"urich, 2007 [arXiv:math-ph/0603038]. 

 \bibitem{delapham}
E. Delabaere and F. Pham, ``Resurgent methods in semi-classical asymptotics," Ann. Inst. Henri Poincar\'e {\bf 71} (1999) 1.   
 


\bibitem{dfgzj}
P.~Di Francesco, P.~Ginsparg and J.~Zinn-Justin, ``2-D Gravity and random matrices,''
Phys.\ Rept.\  {\bf 254}, 1 (1995)
[arXiv:hep-th/9306153].

\bibitem{dv}
R.~Dijkgraaf and C.~Vafa,
``Matrix models, topological strings, and supersymmetric gauge theories,''
Nucl.\ Phys.\ B {\bf 644}, 3 (2002)
[arXiv:hep-th/0206255].

\bibitem{ds}
M.~R.~Douglas and S.~H.~Shenker, ``Strings In Less Than One-Dimension,''
Nucl.\ Phys.\ B {\bf 335}, 635 (1990).

\bibitem{ecalle}
J. \'Ecalle, {\it Les fonctions resurgentes}. 

\bibitem{john}
 J.~R.~Ellis, E.~Gardi, M.~Karliner and M.~A.~Samuel, ``Pad\'e Approximants, Borel Transforms and Renormalons: the Bjorken Sum Rule 
 as a Case Study,''
  Phys.\ Lett.\  B {\bf 366}, 268 (1996)
  [arXiv:hep-ph/9509312].
  
\bibitem{eynard}
B.~Eynard, ``Large $N$ expansion of convergent matrix integrals, holomorphic anomalies, and background independence,''
  arXiv:0802.1788 [math-ph].
  
 \bibitem{eynardproof}
 B.~Eynard, ``All orders asymptotic expansion of large partitions,''
  arXiv:0804.0381 [math-ph].
 
 \bibitem{eo}
 B.~Eynard and N.~Orantin, ``Invariants of algebraic curves and topological expansion,''
  arXiv:math-ph/0702045.
 
 \bibitem{ezj}
B.~Eynard and J.~Zinn-Justin, ``Large order behavior of 2-D gravity coupled to $d <1$ matter,''
  Phys.\ Lett.\  B {\bf 302}, 396 (1993)
  [arXiv:hep-th/9301004].
  
 \bibitem{fateev} 
  V.~A.~Fateev, V.~A.~Kazakov and P.~B.~Wiegmann, ``Principal Chiral Field At Large N,''
  Nucl.\ Phys.\  B {\bf 424}, 505 (1994)
  [arXiv:hep-th/9403099].
  
\bibitem{fr} 
G.~Felder and R.~Riser, ``Holomorphic matrix integrals,''
  Nucl.\ Phys.\  B {\bf 691}, 251 (2004)
  [arXiv:hep-th/0401191].

 
 \bibitem{painlevet}
 A.~S.~Fokas, A.~R.~Its, A. A. Kapaev and V. Yu. Novokshenov, {\it Painlev\'e Transcendents: A Riemman-Hilbert Approach}, AMS (2006). 
   
 \bibitem{fik}
  A.~S.~Fokas, A.~R.~Its and A.~V.~Kitaev, ``Discrete Painlev\'e equations and 
  their appearance in quantum gravity,'' Commun.\ Math.\ Phys.\  {\bf 142}, 313 (1991);
``The Isomonodromy Approach To Matrix Models In 2-D Quantum Gravity,''
  Commun.\ Math.\ Phys.\  {\bf 147}, 395 (1992); 

 
 \bibitem{fgs}
V. Franceschini, V. Grecchi and H.J. Silverstone, ``Complex energies from real perturbation series for 
the LoSurdo-Stark effect in hydrogen by Borel-Pad\'e approximants," Phys. Rev. A {\bf 32}, 1338 (1985).

\bibitem{garcia}
S.~Garc\' \i a, Z.~Guralnik and G.~S.~Guralnik, ``Theta vacua and boundary conditions of the Schwinger-Dyson equations,''
  arXiv:hep-th/9612079.

 \bibitem{morozov}
 A.~Gerasimov, A.~Marshakov, A.~Mironov, A.~Morozov and A.~Orlov, ``Matrix models of 2-D gravity and Toda theory,''
  Nucl.\ Phys.\  B {\bf 357}, 565 (1991).
  
\bibitem{lattice}
 Y.~Y.~Goldschmidt, 
  ``1/N Expansion In Two-Dimensional Lattice Gauge Theory,''
  J.\ Math.\ Phys.\  {\bf 21}, 1842 (1980).
  
\bibitem{gma}
 D.~J.~Gross and A.~Matytsin, ``Instanton induced large N phase transitions in two-dimensional and four-dimensional QCD,''
  Nucl.\ Phys.\  B {\bf 429}, 50 (1994)
  [arXiv:hep-th/9404004].
  
 \bibitem{gm}
 D.~J.~Gross and A.~A.~Migdal,
``Nonperturbative Two-Dimensional Quantum Gravity,''
Phys.\ Rev.\ Lett.\  {\bf 64}, 127 (1990).
 
\bibitem{gw}
 D.~J.~Gross and E.~Witten,
  ``Possible Third Order Phase Transition In The Large N Lattice Gauge
  Theory,''
  Phys.\ Rev.\ D {\bf 21}, 446 (1980).

\bibitem{grunberg} 
 G.~Grunberg, ``Perturbation theory and condensates,''
  Phys.\ Lett.\  B {\bf 325}, 441 (1994).
  
\bibitem{guralnik}
G.~Guralnik and Z.~Guralnik, ``Complexified Path Integrals and the Phases of Quantum Field Theory,''
  arXiv:0710.1256 [hep-th].
 \bibitem{hoy}
  N.~Halmagyi, T.~Okuda and V.~Yasnov, ``Large N duality, lens spaces and the Chern-Simons matrix model,''
  JHEP {\bf 0404}, 014 (2004)
  [arXiv:hep-th/0312145].
  
  \bibitem{hy}
  N.~Halmagyi and V.~Yasnov, ``The spectral curve of the lens space matrix model,''
  arXiv:hep-th/0311117.

  
\bibitem{lvm}
M.~Hanada, M.~Hayakawa, N.~Ishibashi, H.~Kawai, T.~Kuroki, Y.~Matsuo and T.~Tada, ``Loops versus matrices: The nonperturbative aspects of noncritical string,''
  Prog.\ Theor.\ Phys.\  {\bf 112}, 131 (2004)
  [arXiv:hep-th/0405076].

\bibitem{hm}
S.P. Hastings and J.B. McLeod, ``A boundary value problem associated with the second Painlev\'e transcendent and the Korteweg-de Vries equation," Arch. Ration. Mech. Anal. {\bf 73} (1980) 31.

\bibitem{iy}
N.~Ishibashi and A.~Yamaguchi, ``On the chemical potential of D-instantons in c = 0 noncritical string theory,''
  JHEP {\bf 0506}, 082 (2005)
  [arXiv:hep-th/0503199].

\bibitem{kapaev}
A.R. Its and A.A. Kapaev, ``Quasi-linear Stokes phenomenon for the second Painlev\'e transcendent," Nonlinearity {\bf 16}, 363 (2003)[arXiv:nlin.SI/0108010].

 \bibitem{jentschura}
U.~D.~Jentschura, ``The resummation of nonalternating divergent perturbative expansions,''
  Phys.\ Rev.\  D {\bf 62}, 076001 (2000)
  [arXiv:hep-ph/0001135].
 
\bibitem{jz}
 J.~Jurkiewicz and K.~Zalewski,
 ``Vacuum Structure Of The U(N $\to$ Infinity) Gauge Theory On A Two-Dimensional Lattice For A Broad Class Of Variant Actions,''
  Nucl.\ Phys.\ B {\bf 220}, 167 (1983). 
  
  
\bibitem{kawai}
H.~Kawai, T.~Kuroki and Y.~Matsuo, 
``Universality of nonperturbative effect in type 0 string theory,''
  Nucl.\ Phys.\  B {\bf 711}, 253 (2005)
  [arXiv:hep-th/0412004].

\bibitem{kms}
I.~R.~Klebanov, J.~M.~Maldacena and N.~Seiberg, ``Unitary and complex matrix models as 1-d type 0 strings,''
  Commun.\ Math.\ Phys.\  {\bf 252}, 275 (2004)
  [arXiv:hep-th/0309168].

\bibitem{kmt}
A.~Klemm, M.~Mari\~no and S.~Theisen,
``Gravitational corrections in supersymmetric gauge theory and matrix models,''
JHEP {\bf 0303}, 051 (2003)
[arXiv:hep-th/0211216].

\bibitem{mmss}
 J.~M.~Maldacena, G.~W.~Moore, N.~Seiberg and D.~Shih, ``Exact vs. semiclassical target space of the minimal string,''
  JHEP {\bf 0410}, 020 (2004)
  [arXiv:hep-th/0408039].
  
  \bibitem{mandal}
  G.~Mandal, ``Phase Structure Of Unitary Matrix Models,''
  Mod.\ Phys.\ Lett.\ A {\bf 5}, 1147 (1990).

 \bibitem{mmint}
 M.~Mari\~no,
  ``Chern-Simons theory, matrix integrals, and perturbative three-manifold invariants,''
  Commun.\ Math.\ Phys.\  {\bf 253}, 25 (2004)
  [arXiv:hep-th/0207096].
  
\bibitem{mmleshouches}
 M.~Mari\~no, ``Les Houches lectures on matrix models and topological strings,''
  arXiv:hep-th/0410165.
  
\bibitem{mm}
M.~Mari\~no, ``Open string amplitudes and large order behavior in topological string theory,''
  JHEP {\bf 0803}, 060 (2008)
  [arXiv:hep-th/0612127].
 
\bibitem{msw}
M.~Mari\~no, R.~Schiappa and M.~Weiss, ``Nonperturbative Effects and the Large-Order Behavior of Matrix Models and Topological Strings,''
  arXiv:0711.1954 [hep-th].
  
\bibitem{mswta}
M.~Mari\~no, R.~Schiappa and M.~Weiss, ``Multi-Instantons and Multi-Cuts,''
  arXiv:0809.2619 [hep-th].

\bibitem{mizoguchi}
S.~Mizoguchi, ``On unitary/Hermitian duality in matrix models,''
  Nucl.\ Phys.\  B {\bf 716}, 462 (2005)
  [arXiv:hep-th/0411049].
 
 
\bibitem{neuberger}
 H.~Neuberger, ``Nonperturbative Contributions In Models With A Nonanalytic Behavior At Infinite N,''
  Nucl.\ Phys.\  B {\bf 179}, 253 (1981).

\bibitem{p}  
R. Pandharipande, ``The Toda Equation and the Gromov--Witten Theory of the Riemann Sphere," Lett.\ Math.\ Phys.\ \textbf{53} (2000) 59 [arXiv:math.AG/9912166].

\bibitem{ps}
V.~Periwal and D.~Shevitz, ``Unitary Matrix Models As Exactly Solvable String Theories,''
  Phys.\ Rev.\ Lett.\  {\bf 64}, 1326 (1990); ``Exactly Solvable Unitary Matrix Models: Multicritical Potentials And Correlations,''
  Nucl.\ Phys.\ B {\bf 344}, 731 (1990).
  
\bibitem{pskpz}
M. Prah\"ofer and H. Spohn, ``Exact scaling functions for one-dimensional stationary KPZ growth," J. Stat. Phys. {\bf 115}, 255 (2004) 
[arXiv:cond-mat/0212519].

\bibitem{raczka}
P.~A.~Raczka, ``Summation Of The Perturbation Expansion And The QCD Corrections To The Tau Decay,''
  Phys.\ Rev.\  D {\bf 43}, 9 (1991).

\bibitem{rv} 
P.~Rossi, M.~Campostrini and E.~Vicari, ``The large-N expansion of unitary-matrix models,'' Phys.\ Rept.\  {\bf 302}, 143 (1998)
  [arXiv:hep-lat/9609003].

\bibitem{seara}
T. M. Seara and D. Sauzin, ``Ressumaci\'o de Borel i teoria de la ressurg\`encia," 
Butl. Soc. Catalana Mat. {\bf 18} (2003) 131. 

\bibitem{ss}
 N.~Seiberg and D.~Shih, ``Flux vacua and branes of the minimal superstring,''
  JHEP {\bf 0501}, 055 (2005)
  [arXiv:hep-th/0412315].
  
\bibitem{shenker} 
S.H.~Shenker, ``The strength of nonperturbative effects in string theory," 
in O.~\'Alvarez, E.~Marinari and P.~Windey (eds.), {\it Random Surfaces and Quantum Gravity}, Plenum, New York 1992.

 \bibitem{sy}
 P.~G.~Silvestrov and A.~S.~Yelkhovsky, ``Two-dimensional gravity as analytical continuation of the random matrix 
 model,''
  Phys.\ Lett.\  B {\bf 251}, 525 (1990).
  
 \bibitem{tw} 
  C.~A.~Tracy and H.~Widom, ``Level spacing distributions and the Airy kernel,''
  Commun.\ Math.\ Phys.\  {\bf 159}, 151 (1994)
  [arXiv:hep-th/9211141].
  
\bibitem{wadia}
 S.~R.~Wadia, ``A Study Of U(N) Lattice Gauge Theory In Two-Dimensions,'' EFI-79/44-CHICAGO, Jul. 1979.
  
\bibitem{qmwadia}
S.~R.~Wadia, ``N = Infinity Phase Transition In A Class Of Exactly Soluble Model Lattice Gauge Theories,''
  Phys.\ Lett.\  B {\bf 93}, 403 (1980).

\bibitem{zj}
J. Zinn--Justin, {\it Quantum field theory and critical phenomena}, Oxford University Press, 2002.
\bibitem{zjj}
J.~Zinn-Justin and U.~D.~Jentschura,  ``Multi-instantons and exact results I: Conjectures, WKB expansions, and instanton interactions,''
  Annals Phys.\  {\bf 313}, 197 (2004) [arXiv:quant-ph/0501136];
  ``Multi-Instantons And Exact Results II: Specific Cases, Higher-Order Effects, And Numerical Calculations,''
  Annals Phys.\  {\bf 313}, 269 (2004)
  [arXiv:quant-ph/0501137].

 
  \end{thebibliography}

\end{document}